\documentclass[aps,prd,twocolumn,superscriptaddress,floatfix,nofootinbib]{revtex4}

\usepackage{color}
\usepackage{graphicx}

\newcommand{\lsim}{\mathrel{\mathop{\kern 0pt \rlap
      {\raise.2ex\hbox{$<$}}}\lower.9ex\hbox{\kern-.190em $ \sim$}}}
\newcommand{\gsim}{\mathrel{\mathop{\kern 0pt
      \rlap{\raise.2ex\hbox{$>$}}}\lower.9ex\hbox{\kern-.190em $\sim$}}}
\newcommand{\beq}{\begin{equation}}
\newcommand{\eeq}{\end{equation}}

\newcommand{\be}{\begin{equation}}
\newcommand{\ee}{\end{equation}}
\newcommand{\beqarr}{\begin{eqnarray}}
\newcommand{\eeqarr}{\end{eqnarray}}

\newcommand{\sigmap}{\sigma^p_{\phi\gamma}}

\begin{document}

\title{Long--Range Forces in Direct Dark Matter Searches }
\thanks{Preprint numbers: CERN-PH-TH/2011-174 and DFTT 12/2011}

\author{N. Fornengo}
\affiliation{Dipartimento di Fisica Teorica, Universit\`a di Torino, I-10125 Torino, Italy}
\affiliation{Istituto Nazionale di Fisica Nucleare, Sezione di Torino I-10125 Torino, Italy}
\author{P. Panci}
\affiliation{CERN Theory Division, CH-1211 Gen\`eve, Switzerland}
\affiliation{Consorzio Interuniversitario per la Fisica Spaziale (CIFS), I-10133 Torino, Italy}
\affiliation{Istituto Nazionale di Fisica Nucleare, Laboratori Nazionali del Gran Sasso, 67010 Assergi (AQ), Italy}
\author{M. Regis}
\affiliation{Dipartimento di Fisica Teorica, Universit\`a di Torino, I-10125 Torino, Italy}
\affiliation{Istituto Nazionale di Fisica Nucleare, Sezione di Torino I-10125 Torino, Italy}
%

%

\date{\today}

\begin{abstract}

We discuss the positive indications of a possible dark matter signal in direct detection experiments in terms of a mechanism of interaction between the dark matter particle and the nuclei occurring via the exchange of a light mediator, resulting in a long--range interaction.  We analyze the annual modulation results observed by the DAMA and CoGeNT experiments and  the observed excess of events of CRESST. In our analysis, we discuss the relevance of uncertainties related to the velocity distribution of galactic dark matter and to the channeling effect in NaI. We find that a long--range force is a viable mechanism, which can provide full agreement between the reconstructed dark matter properties from the various experimental data sets, especially for masses of the light mediator in the 10--30 MeV range and a light dark matter with a mass around 10 GeV. The relevant bounds on the light mediator mass and scattering cross section are then derived, should the annual modulation effects be due to this class of long--range forces.

\end{abstract}

\pacs{95.35.+d,11.30.Pb,12.60.Jv,95.30.Cq}

\maketitle

\section{Introduction}

Dark--matter direct--detection experiments have been providing exciting results in terms of 
measured features which have the right properties to be potentially ascribed to a dark matter (DM) signal. The typical effect of annual modulation of the recoil
rate \cite{freese} has been put under deep scrutiny by the DAMA/NaI Collaboration 
starting more than a decade ago \cite{dama1997}. Annual modulation of the rate with viable DM interpretation was observed. 
The upgraded DAMA/LIBRA detector has confirmed \cite{dama2008}, with much larger statistics, the
annual modulation effect, reaching the unprecedented result of an evidence of 8.9 $\sigma$ C.L. for the cumulative exposure \cite{dama2010}. The DAMA annual modulation effect has been shown to be
compatible with a DM effect which, for the case of a coherent scattering, refer to
a range of DM masses which spans
from a few GeV up to a few hundred of GeVs and cross sections between
$10^{-42}$ cm$^2$ to $10^{-39}$ cm$^2$ \cite{dama1997,dama2008,dama2010}, and with some noticeable differences due to the galactic halo modeling \cite{Belli:2002yt,Belli:2011kw}.

More recently, the CoGeNT experiment first reported an irreducible excess in their
counting rate \cite{cogentrate}, which could be in principle ascribed to a DM signal.
In the last months, the same experiment reported an additional analysis which shows
that the time--series of their rate is actually compatible with an annual modulation
effect \cite{cogentmod}. The evidence of modulation in CoGeNT is at the level of 2.8 $\sigma$ C.L..

The interesting feature is that the DAMA and CoGeNT results appear to be compatible
for relatively light DM particles, in the few GeV to tens of GeV mass range 
and coherent scattering cross section around $10^{-41}$ cm$^2$ to $10^{-40}$ cm$^2$
\cite{Belli:2011kw} (as usual, the actual relevant range of masses and cross section depends on
the assumptions on the galactic DM properties, namely the velocity distribution
function and the local DM density \cite{Belli:2011kw}). Further relevant analyses
can be found in Refs. \cite{Foot:2011pi,Schwetz:2011xm,Farina:2011pw,McCabe:2011sr,Fox:2011px} 
and Refs. \cite{Hooper:2011hd,Gondolo:2011eq,DelNobile:2011je,Arina:2011si,Frandsen:2011ts,Kaplan:2011yj,Feng:2011vu,Fitzpatrick:2010br,Hooper:2010uy,Foot:2010rj,Chang:2010yk,Fitzpatrick:2010em,Kopp:2009qt}.

In this paper we discuss an alternative possibility, namely the case of a DM particle that scatters on the nucleus with long--range interactions, like those that are induced by a light mediator, such that the nature of the DM--nucleus cross interaction is not contact--like. An example of these kind of interactions is given by mirror photons in models of mirror dark matter 
\cite{Blinnikov:1982eh, Blinnikov:1983gh, Foot:1991bp, Foot, Foot:2011pi, Berezhiani, Berezhiani:2005ek, Ciarcelluti:2010zz}. We analyze if and under what circumstances a long--range
force can explain the positive hints of a signal in direct detection experiments and
what kind of bounds could be derived by those evidences on the light mediator mass
and scattering cross--section (or, alternatively, coupling constants). 

We discuss the impact of the light--mediator parameters (mostly its mass, which determines
the level of deviation from the standard case of a contact--like scattering) on the reconstruction of the DM mass. We show that long--range forces
mediated by a 10--30 MeV boson may provide compatibility between the different experimental
direct--detection results. These results are discussed for some variation of the galactic halo models.

We concentrate our analysis mostly on the DAMA and CoGeNT results, since they are based
on a specific feature of the DM signal, namely the annual modulation, which is hardly
mimicked with the correct features (period, phase, energy range,
size) by background sources. However we will discuss also the relevance of the CRESST irreducible excess of events \cite{Jochum:2011zz,Angloher:2011uu}, which is currently based on the total event rate (without resorting to time dependence)
but nevertheless points toward an additional indication of a possible signal.

CDMS and XENON experiments have recently reported a small number of events which pass all the selection cuts (2 events for CDMS \cite{Ahmed:2009zw} and 6 events
for XENON 100, reduced to 3 events after post--selection analysis \cite{xenon100}), 
still too few to be correlated to a signal. They therefore can provide upper bounds on the DM scattering cross--section. The actual response of these detector to the light DM
case which is here under scrutiny has been critically analyzed and appears to be uncertain and model dependent for light DM \cite{dama/xenon,collar/xenon,collar/cdms}. We will show the results obtained from CDMS and XENON in our analysis on long--range forces, but we will not enforce those bound.

\section{Direct detection signals}
\label{sec:signals}

Direct detection relies on the direct scattering of DM particles off the nuclei
of ordinary matter, the two main processes being elastic scattering:
\beq
\chi+ {\cal N}(A,Z)_{\rm at \,rest}\rightarrow  \chi+ {\cal N}(A,Z)_{\rm recoil},
\label{eq:elastic}
\eeq
and inelastic scattering:
\beq
\chi+ {\cal N}(A,Z)_{\rm at \,rest}\rightarrow  \chi'+ {\cal N}(A,Z)_{\rm recoil},
\label{eq:inelastic}
\eeq
In Eqs. (\ref{eq:elastic}) and (\ref{eq:inelastic}) $\chi$ and $\chi'$ are the dark matter particles and its excited state and $A$, $Z$ are the atomic mass and atomic number of nucleus 
$\cal N$, respectively. In the detector rest frame, a DM particle
with velocity $v$ and mass $m_\chi$, would produce a nuclear recoil of energy $E_R$.
The minimal velocity providing a nuclear recoil energy $E_{R}$ is:
\beq
\label{minvelocity}
v_{\rm min}(E_{R})=\sqrt{ \frac{m_{N}\,E_{R}}{2\mu_{\chi {N}}^2} }\left(1+\frac{\mu_{\chi {N}}\,\delta}{m_{N}\,E_{R}}\right)
\eeq
where $\delta=m_\chi'-m_\chi$ the mass splitting between $\chi$ and $\chi'$. Elastic scattering occurs for $\delta = 0$, while $\delta \neq 0$ implies inelastic scattering.

In this paper we will consider only the case of elastic scattering, but we will extend the
mechanism of interaction to the possibility that a light mediator may induce a long--range
interaction, instead of the typical situation where the scattering cross--section is obtained
through a contact interaction, like it is the case for the exchange of heavy bosons.
Long range forces alter the detector response to DM interaction, as will be outlined
in the remainder of the Section, and can make low--threshold detectors (like CoGeNT and DAMA) especially sensitive to
the presence of these long--range forces.

\subsection{Generalization of the point-like cross section to long--range interactions}

Long range interactions can be described by means of a light (massless in the
extreme limit) mediator $\phi$ which, in
the non--relativistic limit, suitable for the DM--nucleus scattering 
in DM direct detection, corresponds to the presence of a Yukawa potential, whose scale
is determined by the mass $m_\phi$ of the mediator. A specific realization is offered by models of mirror dark matter and models where mirror photons possess a kinetic mixing with ordinary photons \cite{Blinnikov:1982eh, Blinnikov:1983gh, Foot:1991bp, Foot, Foot:2011pi, Berezhiani, Berezhiani:2005ek, Ciarcelluti:2010zz, Mambrini:2011dw, McDonald:2010fe, Bullimore:2010aj, Mambrini:2010dq, Chun:2010ve, Gaete:2010tm, Batell:2009vb, Cheung:2009qd, Abel:2008ai, Foot:2000vy, Dedes:2009bk, Pospelov:2008jd, Pospelov:2007mp, Antoniadis:2002cs, Holdom:1985ag}. In this case, mirror charged particle couples to ordinary nucleus with electric charge $Ze$ ($Z$ being the number of protons in the nucleus), with an effective coupling  $\epsilon \,Z' g_{\rm dark}$ (being $Z' g_{\rm dark}$ the coupling between DM and mirror photon, and with $\epsilon$ parametrizing the kinetic mixing between mirror and ordinary photon). 
The radial dependent Yukawa potential of the interaction can be cast in the form:

\beq
V(r)=\left(\alpha_{\rm SM} \alpha_{\rm dark}\right)^{\frac12}\frac{\epsilon \,Z\,Z'}r\,e^{-m_\phi r},
\label{YukPot}
\eeq
where $\alpha_{\rm SM} = e^2/(4\pi)$ is the electromagnetic fine structure constant and
$\alpha_{\rm dark} = g_{\rm dark}^2/(4\pi)$. In a more general framework, one replaces
$[(\alpha_{\rm SM} \alpha_{\rm dark})^{\frac12} \epsilon \,Z\, Z']$ with the relevant 
coupling factors between DM and the nucleus, which may be just on protons, or on protons and
neutrons with suitable strengths determined by the specific model. For definiteness,
we consider here the case motivated by mirror photons, which implies interactions with
protons only.

From the potential in Eq. (\ref{YukPot}), one obtains the differential cross section:

\beq
\label{dsigmadEr}
\frac{d\sigma(v,E_{R})}{dq^2}=\frac{2m_N\lambda}{\left(q^2+m_\phi^2\right)^2}\frac1{v^2}
\, F^2(E_R) \,, 
\eeq
where $q^2=2 m_N E_{R}$ is the square of the momentum transferred in the interaction,
$v$ is the speed of the DM particle, $F(E_R)$ denotes the nuclear form factor which takes into account the finite dimension of the nucleus and:

\beq
\lambda=\frac{2\pi\alpha_{\rm SM}\alpha_{\rm dark}\epsilon^2Z^2 Z'^2}{m_{ N}}\,.
\eeq
The differential cross section of Eq. (\ref{dsigmadEr}) exhibits two limits. The point--like limit of the interaction occurs when the mass of the mediator is much larger than the 
transferred momentum, {\em i.e.} when $m_\phi \gg q$. By ``point--like" in this context we
mean that the mechanism of interactions is realized through a contact interaction.
In this regime the differential cross section reduces to the standard case:

\beq
\frac{d\sigma(v,E_{R})}{dE_{R}}=\frac{m_N}{2 \mu_{\chi \rm p}^2}\frac1{v^2}\,Z^2\sigma_{\phi\gamma} \, F^2(E_R)\, , 
\label{eq:diffrate}
\eeq
where $\mu_{\chi \rm p}$ is the DM--nucleon reduced mass and
the total point--like cross section per nucleon for such coupling is given by:
\begin{equation}\label{TotalCrossSection}
\sigma_{\phi\gamma}=\frac{16\pi \epsilon^2 Z'^2\alpha_{\rm SM}\alpha_{\rm dark}}{m_\phi^4}\mu_{\chi\rm p}^2.
\end{equation}
Eqs. (\ref{eq:diffrate},\ref{TotalCrossSection}) can be generalized to the case of scattering off both protons and
nucleons with a change such that
 $ Z^2 \rightarrow  [Z + f_n/f_p (A-Z)]^2$,
where $A$ is the mass
number of the nucleus and $f_{p,n}$ are factors which differentiate
the coupling on protons and neutrons. The case we are considering, motivated by a mirror--photon exchange, refers to $f_n = 0$. The allowed
regions and bounds we will derive in the plane $\sigmap$ -- $m_{\chi}$ will therefore reflect this fact, and in particular will be shifted with respect to the standard case
$f_p=f_n$ which usually arises for many DM candidates. For analyses which relax the
assumption $f_p=f_n$ see e.g. \cite{Mambrini:2010dq,Gondolo:2011eq,DelNobile:2011je,Gao:2011ka,Feng:2011vu,Pato:2011de,Chen:2011vd,Frandsen:2011cg,Kang:2011wb,Kang:2010mh,Kurylov:2003ra,Cline:2011zr}.

The long--range nature of the interaction occurs when $m_\phi \ll q$.
In this regime the differential cross section acquires an explicit dependence on the nuclear recoil energy and a Rutherford--like cross section emerges:

\beq
\frac{d\sigma(v,E_{R})}{dE_{R}}=
\frac\lambda{E_{\rm R}^2}\frac1{v^2}\, F^2(E_R)\,
\propto \, E_{R}^{-2},
\eeq

The $E_{R}^{-2}$ drop--off of the cross--section, which is not present in the point--like case, 
makes experiments with low energy thresholds (like DAMA and CoGeNT) to respond better 
to the interaction mechanism, as compared to experiments with relatively
high energy threshold (CMDS and XENON100 have stable thresholds of the order of/larger than 10 keV),
and so may in principle improve the compatibility among those experiments.
The recoil energy at which the interaction becomes effectively long--range depends on the
mass of the target. In particular, larger is the mass smaller is such transition energy.
Therefore, for intermediate mass of the dark photon, a possible feature may arise, such that,
at a given recoil energy, the interaction might be effectively long--range in experiments 
with large target--mass, while being in the point--like limit in low target--mass experiments.

Considering typical a mass of targets $m_N\sim$ 100 GeV and nuclear recoil energy windows around few to tens of keV, the long range nature of the interaction manifests itself if the mass of the dark photon is lesser than 10 MeV.  

More generally, the differential cross section of Eq. (\ref{dsigmadEr}) in terms of a normalized total cross section $\sigmap=\sigma_{\phi\gamma}(m_\phi\equiv \tilde m_\phi)$, is: 
%
\beq
\frac{d\sigma(v,E_{R})}{dE_{R}}=\frac{m_N}{2 \mu_{\chi \rm p}^2}\frac1{v^2}\,Z^2 \sigmap\left(\frac{\tilde m_\phi^2/(2m_N)}{E_{R}+m_\phi^2/(2m_N)}\right)^2  F^2(E_R),
\eeq
where $\tilde m_\phi=1$ GeV $\simeq m_{\rm p}$. We will use as free parameters in our analysis $\sigmap$ and $m_{\phi}$, an alternative choice being $m_{\phi}$ and the effective coupling of the DM with the nucleon 
$\epsilon g_{\rm dark}$.

As for the nuclear form factors, we adopt the standard form described by Helm in Ref. \cite{Helm}:
\beq
\label{NuclFormSI1}
F(q r_N)=3\,\frac{j_1(q r_N)}{q r_N}\exp[-(q\,s)^2/2]
\eeq
where $j_1$ is the spherical Bessel function of the first kind with $n=1$:
\beq
\label{NuclFormSI2}
j_1(q r_N)=\frac{\sin (q r_N)}{(q r_N)^2}-\frac{\cos (q r_N)}{(q r_N)},
\eeq
In Eqs. (\ref{NuclFormSI1},\ref{NuclFormSI2}) $r_N$ is the nuclear radius and $s$ is estimated trying to reproduce the more accurate results obtained from numerical evaluation of the Fourier transform relative to a Fermi distribution of scattering centers. A good agreement is obtained for $s \simeq (197 \,\mbox{MeV})^{-1}$ and 
$r_N = ((1/(164 {\rm MeV})A^{1/3})^2-5s^2)^{1/2}$. We remind that this expression of spin--independent form factor is derived assuming a Fermi distribution for the nuclear charge and that all the parameters used in this parameterization may be affected by sizable uncertainties.

\subsection{Rate of nuclear recoils}
\begin{figure}[t]
\includegraphics[width=0.49\textwidth]{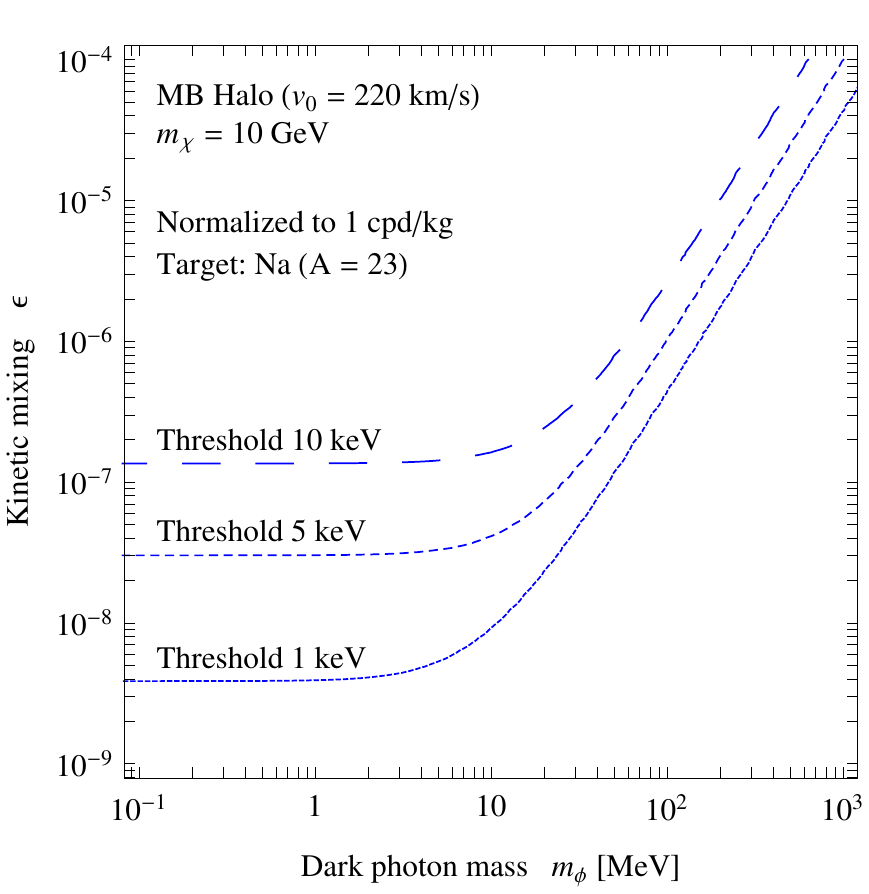}
\caption{In the plane $\epsilon$ vs. $m_\phi$, iso--contours 
of constant rate (chosen as 1 cpd/kg) on a Na target, for a 10 GeV
DM particle scattering and for various values for the energy threshold
are shown. The galactic halo model is an isothermal sphere with Maxwell--Boltzmann
(MB) velocity distribution with velocity dispersion 
$v_0=220$ km s$^{-1}$ and local density
$\rho_0 = 0.3$ GeV cm$^{-3}$. 
}
\label{fig:A}
\end{figure}

The differential recoil rate of a detector can be defined as:

\beq\label{GeneralRate}
\frac{dR}{dE_{R}}=N_T \int \frac{d\sigma(v,E_{R})}{dE_{R}}\, v \, dn_\chi, 
\eeq
where $N_{T}$ is the total number of targets in the detector ($N_A$ is the Avogadro's number) and $dn_\chi$ is the local number density of DM particles with velocities in the elemental volume $d^3v$ around $\vec v$. This last term can be expressed as function of the DM velocity distribution $f_{E}(\vec v)$ in the Earth frame, which is related to the DM 
velocity distribution in the galactic frame $f_{G}(\vec w)$ by a galilean velocity
transformation as $f_E(\vec v) = f_G(\vec v + \vec v_E(t))$ where $\vec v_E (t)$ is the time--dependent Earth (or detector) velocity with respect to the galactic frame. The prominent
time--dependence (on the time--scale of an experiment)
comes from the annual rotation of the Earth around the Sun, which is the origin of the
annual modulation effect of the direct detection rate \cite{freese}. More specifically:

\beq
\vec v_E(t) = \vec v_G + \vec v_S + \vec v_\oplus(t)
\eeq
The galactic rotational velocity of our local system $\vec v_G$ and the Sun's proper motion
$\vec v_S$ are basically aligned and their absolute values are $v_G \equiv v_0 = 220 \pm 50$ km s$^{-1}$ and $v_S = 12$ km s$^{-1}$, while the Earth rotational velocity $\vec v_\oplus(t)$ has
a size $v_\oplus = 30$ km s$^{-1}$, period of 1 year and phase such that it is aligned to
$\vec v_G$ around June 2nd and it is inclined of an angle $\gamma \simeq 60$ degrees with
respect to the galactic plane. More details can be found, for instance, in Ref. \cite{distortion}.
Summarizing:
\beq\label{densitynumberDD}
dn_\chi=n_\chi f_{\rm E}(\vec v)\,d^3v,
\eeq
where $n_\chi=\xi_\chi\rho_0/m_\chi$ is the local DM number density in the Galaxy
and is determined by the local dark DM matter density $\rho_0$ and, in general,
on a scaling factor $\xi$ which accounts for the possibility that
the specific DM candidate under consideration does not represent the whole amount of DM.
For all practical purposes, we can just assume $\xi=1$ here. In Eq. 
(\ref{densitynumberDD}) the velocity distribution function needs to be properly normalized. This
can be achieved by requiring that in the galactic frame:

\beq
\int_{v\leq v_{\rm esc}} d^3v \,f_G(\vec v) = 1
\eeq
where $v_{\rm esc}$ denotes the escape velocity of DM particles 
in the Milky Way. For definiteness, we will adopt here $v_{\rm esc} = 650$ km s$^{-1}$.

When considering the generalization of the differential cross--section to the case of
long--range forces, the differential rate of nuclear recoil can be cast in the form:

\beq\label{FactorizedRate}
\frac{dR}{dE_{R}}(t) = 
N_0 \, \frac{\xi_\chi \rho_0}{m_\chi} \,
\frac{m_{N}}{2\mu^2_{\chi p}} \,\, (Z^2 \sigmap) \,\, {\mathcal I(v_{\rm min},t)} \,\,
\mathcal G(E_R)
\eeq
where:
\beq
{\cal G}(E_R) = \left(\frac{\tilde m_\phi^2/(2m_N)}{E_{R}+m_\phi^2/(2m_N)}\right)^2 \, F^2(E_R)
\eeq
and:
\beq
{\mathcal I(v_{\rm min},t)} = \int_{v \geq v_{\rm min(E_R)}} \frac{f_E(\vec v)}{v}\, d^3 v
\eeq
with $v_{\rm min} (E_R)$ given in Eq. (\ref{minvelocity}).
The detection rate is function of time through the function $\mathcal I(v_{\rm min},t)$,
as a consequence of the annual motion of the Earth around the Sun. The actual form of
the function $\mathcal I(v_{\rm min},t)$ depends on the velocity distribution function
of the DM particles in the halo. We will consider two cases: an isothermal sphere, whose
velocity distribution function in the galactic frame $f_G({\vec v})$ is a Maxwell-Boltzmann
function, and a triaxial model, with an anisotropic $f_G({\vec v})$. We will discuss in more
details our choices in the next Section.

Since the Earth velocity $\vec v_{\rm E}(t)$ has an explicit dependence on time due to the movement of the Earth around the Sun, and since this last velocity component is relatively
small when compared to the main boost component represented by $\vec v_G + \vec v_S$, it
is convenient to define a time--dependent parameter $\eta(t)$ as:
\beq\label{vEarth}
\eta_{\rm E}(t)=\eta_\odot+\Delta\eta \, \cos\left[2\pi(t-\phi)/\tau\right] 
\eeq 
where $\eta_\odot = (v_G+v_S)/v_0$ and $\Delta \eta = v_\oplus\cos\gamma/v_0$, with $\Delta\eta \ll \eta_\odot$, and where $\phi = 152.5$ days (June 2nd) is the phase and $\tau=365$ days is the
period of the Earth motion around the Sun. In Eq. (\ref{vEarth}) the time $t$ is clearly
expressed in days. By means of the approximation in Eq. (\ref{vEarth}) we can define a convenient expansion of the recoil rate, which is suitable for velocity distribution functions
which are not strongly anisotropic:

\begin{eqnarray}
&&\frac{dR}{dE_R}(t) \simeq \label{firstTaylorRate} \\
&& \left.\frac{dR}{dE_R}\right|_{\eta_{\rm E}=\eta_\odot}
+\frac\partial{\partial\eta_{\rm E}}\left.\frac{dR}{dE_{R}}\right|_{\eta_{\rm E}=\eta_\odot}
\Delta\eta \cos\left[2\pi(t-\phi)/\tau\right]. \nonumber
\end{eqnarray}

In order to properly reproduce the experimental recoil rate, we should account for effects
associated with the detector response. We therefore need to include both the effect of partial
recollection of the released energy (quenching) and the energy resolution of the detector.
This can be done by the following energy--transformation and convolution:

\beq
\frac{dR}{dE_{\rm det}}(E_{\rm det})=\int dE'\,\mathcal{K}(E_{\rm det},E')\sum_{i}\frac{dR_i}{dE_{R}}\left(E_{R}=\frac{E'}{q_i}\right),
\label{eq:recoil}
\eeq
where the index $i$ runs over the different nuclear species of the detector,
$E_{\rm det}$ is the detected energy and $q_i$ are the quenching factors of each of the
nuclear species. The function $\mathcal{K}(E_{\rm det},E')$ takes into account the response and energy resolution of the detector, for which we assume, as is generally done, a gaussian behavior. 

As a final step, we need to average the recoil rate of Eq. (\ref{eq:recoil}) over the
energy bins of the detector. We therefore define the unmodulated components of the
rate $S_{0k}$ and the modulation amplitudes $S_{mk}$ for each energy bin $k$ of width
$\Delta E_k$ as:

\beq\label{ExpTotalRateTay1}
S_{0k}=\frac1{\Delta E_k}\int_{\Delta E_k}dE_{\rm det}\,\left.\frac{dR}{dE_{\rm det}}\right|_{\eta_{\rm E}=\eta_\odot}
\eeq

\beq\label{ExpTotalRateTay2}
S_{mk}=\frac1{\Delta E_k}\int_{\Delta E_k}dE_{\rm det}\,\,\,
\left.
\frac\partial{\partial\eta_{\rm E}}\frac{dR}{dE_{\rm det}}
\right|_{\eta_{\rm E}=\eta_\odot}\Delta\eta.
\eeq
$S_{0k}$ and $S_{mk}$ are the relevant quantities we will use in the analysis of the experimental data of DAMA and CoGeNT. In the case of experiments which do not address the annual modulation effect, only $S_{0k}$ are relevant.

\section{Datasets and Analysis technique}
\label{sec:analysis}

  \begin{table}[t]
  \begin{center}
  \begin{tabular}{|l|c|c|}
  \multicolumn{2}{l}{\footnotesize Point-like ($m_\phi=1$ GeV)} \\
  \hline
  & DAMA ChF & CoG Mod \\
  \hline
  \hline
  MB Halo ($v_0=170$ km/s) & 9.56 $\sigma$ & 1.90 $\sigma$ \\
  \hline
  MB Halo ($v_0=220$ km/s) & 9.57 $\sigma$ & 1.77 $\sigma$  \\
  \hline
  MB Halo ($v_0=270$ km/s) & 8.67 $\sigma$ & 1.46 $\sigma$ \\
  \hline
  Triaxial Halo & 9.55 $\sigma$ & 1.82 $\sigma$  \\
  \hline
  \end{tabular}
  \end{center}
  \caption{
 Statistical evidence of presence of modulation for the DAMA and CoGeNT data sets in the case of a point--like interaction ($m_\phi=1$ GeV). 
 This table refers to the analysis performed under the assumption of an isothermal sphere with a Maxwell--Boltzmann (MB) velocity distribution function and for the triaxial halo model discussed in the text. 
For the MB case, the results
for three  different values of the dispersion velocity are reported. The analysis for the DAMA experiment refers to the annual modulation data, bounded by the total (unmodulated) rate \cite{dama2008,dama2010,Nozzoli}, with the fraction of channeling  varied in its allowed interval \cite{Bernabei:2007hw}. The CoGeNT analysis considers the fit to the modulation amplitude \cite{cogentmod}, bounded by the total (unmodulated) rate with L--peaks subtracted.}
  \label{tab1}
  \end{table}

\begin{table}[t]
  \begin{center}
  \begin{tabular}{|l|c|c|}
  \multicolumn{2}{l}{\footnotesize ($m_\phi=10$ MeV)} \\
  \hline
  & DAMA ChF & CoG Mod \\
  \hline
  \hline
  MB Halo ($v_0=170$ km/s) & 9.17 $\sigma$ & 1.69 $\sigma$   \\
  \hline
  MB Halo ($v_0=220$ km/s) & 8.77 $\sigma$ & 1.55 $\sigma$   \\
  \hline
  MB Halo ($v_0=270$ km/s) & 7.74 $\sigma$ & 1.29 $\sigma$   \\
  \hline
  Triaxial Halo & 9.01 $\sigma$ & 1.63 $\sigma$   \\
  \hline
  \end{tabular}
  \end{center}
  \caption{The same as in Tab. \ref{tab1}, for the case of a mediator of mass $m_\phi=10$ MeV.}
  \label{tab2}
  \end{table}

  \begin{table}[t]
  \begin{center}
  \begin{tabular}{|l|c|c|}
  \multicolumn{2}{l}{\footnotesize ($m_\phi=30$ MeV)} \\
  \hline
  & DAMA ChF & CoG Mod \\
  \hline
  \hline
  MB Halo ($v_0=170$ km/s) & 9.50 $\sigma$ & 1.84 $\sigma$   \\
  \hline
  MB Halo ($v_0=220$ km/s) & 9.40 $\sigma$ & 1.69 $\sigma$   \\
  \hline
  MB Halo ($v_0=270$ km/s) & 8.39 $\sigma$ & 1.40 $\sigma$  \\
  \hline
  Triaxial Halo & 9.52 $\sigma$ & 1.77 $\sigma$  \\
  \hline
  \end{tabular}
  \end{center}
  \caption{The same as in Tab. \ref{tab1}, for the case of a mediator of mass $m_\phi=30$ MeV.}
  \label{tab4}
  \end{table}

  \begin{table}[t]
  \begin{center}
  \begin{tabular}{|l|c|c|}
  \multicolumn{2}{l}{\footnotesize Long range ($m_\phi=0$ MeV)} \\
  \hline
  & DAMA ChF & CoG Mod  \\
  \hline
  \hline
  MB Halo ($v_0=170$ km/s) & 8.76 $\sigma$ & 1.61 $\sigma$   \\
  \hline
  MB Halo ($v_0=220$ km/s) & 8.23 $\sigma$ & 1.48 $\sigma$   \\
  \hline
  MB Halo ($v_0=270$ km/s) & 7.27 $\sigma$ & 1.23 $\sigma$   \\
  \hline
  Triaxial Halo & 8.64 $\sigma$ & 1.55 $\sigma$   \\
  \hline
  \end{tabular}
  \end{center}
  \caption{The same as in Tab. \ref{tab1}, for the case of a long--range interaction
 ($m_\phi=0$).}
  \label{tab5}
  \end{table}


Let us now move to the discussion of the techniques we use to analyze the various
data sets. 

For DAMA, CoGeNT  and CRESST, we adopt the same technique of Ref. \cite{Belli:2011kw}:
we test the null hypothesis (absence of modulation for DAMA and CoGeNT, and absence of signal on top of estimated background for CRESST). From this test we obtain
two pieces of information: i) the level at which each data set allows to reject the
null hypothesis (we will find a confidence level of about 8--9 $\sigma$ for DAMA,
1--2 $\sigma$ for CoGeNT, and 4 $\sigma$ for CRESST); ii) in the relevant DM parameter space (defined by the DM mass $m_\chi$ and the DM--proton cross section $\sigma^p_{\phi\gamma}$) we will
determine the domains where
the values of the likelihood--function differ more than $n \sigma$ from the
null hypothesis (absence of modulation), and thus the corresponding evidence of the DM signal.
We will use $n=7,8$, $n=1$, and  $n=3,4$ for DAMA, CoGeNT, and CRESST, respectively \cite{Belli:2011kw}.
This choice (test of the null hypothesis)
allows more proper comparison between the results arising from experimental data sets with different statistical significances and, for the case of DAMA, allows to implement a requirement of a very
high C.L. 

Our statistical estimator is a likelihood function, defined as  
$\mathcal{L}=\prod_i \mathcal{L}_i$, where $i$ stands for the $i$-th energy bin in DAMA and CoGeNT, and for the $i$-th detector in CRESST. $\mathcal{L}_i$ is the likelihood of detecting the number of observed events given the expected background and DM signal.
$\mathcal{L}_i$ are taken to be gaussian for DAMA and CoGeNT and poissonian for CRESST, since in this case the number of events in each sub--detector is low. 
Defining $\mathcal{L}_{bg}$ as the likelihood of absence of signal (i.e., without the DM contribution), the function $\tilde y=-2\ln{\mathcal{L}_{bg}/\mathcal{L}}$ is assumed to be distributed as a $\chi^2$--variable with one degree of freedom, for each value of the DM mass (note that in the DAMA and CoGeNT cases, $\tilde y$ simply reduces to $\tilde y=\chi^2_{bg}-\chi^2$). From the $\tilde y$ function 
we extract, for each value of the DM mass, the
interval on $\sigma^p_{\phi\gamma}$ where the null hypothesis (absence of modulation,
i.e. $\sigma^p_{\phi\gamma} = 0$)
can be excluded at the chosen level of confidence: $7\sigma$ (outer region) or
$8\sigma$ (inner region) for DAMA and $1 \sigma$
for CoGeNT). From this, regions in the $m_\chi$ -- $\sigma^p_{\phi\gamma}$ plane arise.

Constraints from null experiments are derived by constructing again a similar likelihood--function
$\lambda=-2\ln{\mathcal{L}/\mathcal{L}_{bg}}$, where $\mathcal{L}$ is the likelihood of detecting the number of observed events (2 and 3 for CDMS and XENON100, respectively) over the whole energy range of the experiment given the expected background and the DM signal, while in $\mathcal{L}_{bg}$ the DM signal is not included. Both likelihoods are taken to be Poissonian and $\lambda$ is assumed to follow a $\chi^2$-distribution. Bounds are conservatively shown at 5-$\sigma$ C.L.

\subsection{DAMA}
\label{sec:DAMA}
DAMA, located at the Laboratori Nazionali del Gran Sasso, is an observatory for rare processes based on the developments and use of highly radiopure scintillators. The former DAMA/NaI and current DAMA/Libra experiments, made of radiopure NaI(Tl) crystals, have the main aim of investigating the presence of DM particles in the galactic halo by looking at their annual modulation signature. 

The signal in DAMA is the energy deposited in scintillation light. On the other hand, the scattered nucleus is losing energy both electromagnetically and through nuclear interactions: this effect is taken into account by the quenching factors $q$ which convert the total nuclear recoil energy $E_{\rm R}$ to the energy seen by the detector $E_{\rm det} = q E_{\rm R}$. For NaI crystals we take, $q_{\rm Na}$ = 0.3 and $q_{\rm I}$ = 0.09~\cite{Bernabei:1996yh}. Notice that the uncertainty on the actual values of the quenching factors
in NaI \cite {tretyak,Belli:2011kw} can have a visible impact on the reconstructed DM properties \cite{Belli:2011kw}.

However, it has been appreciated \cite{Bernabei:1996yh,Drobyshevski:2008yh} that nuclei recoiling along the characteristic planes of crystals can travel large distances without colliding with other nuclei, and essentially deposit all their energy electromagnetically (which
corresponds to $q = 1$ and thus $E_{\rm det} = E_{\rm R}$).
This process is known as channeling and its relevance in DM direct detection experiments (and in particular for NaI crystals) is currently under scrutiny \cite{Bernabei:2007hw,Bozorgnia:2010xy,Feldstein:2010hw}. Since the actual amount of channeling in
a detector like DAMA is not currently know, we take the following three--fold approach: i)
we show the effect induced by a sizable channeling effect, at the level estimated by the DAMA Collaboration in Ref. \cite{Bernabei:2007hw}, by employing an energy--dependent
channeling fraction $f_{\rm ch}$ as reported in Fig.~4 of~\cite{Bernabei:2007hw}; ii) we consider the case of a negligible channeling effect; iii) due to these uncertainties in the knowledge of the amount of channeling, and considering that the actual value of $f_{\rm ch}$ is likely to lie between the two previous cases, we smoothly vary $f_{\rm ch}$ between them and marginalize over it. We consider this last approach as the most general and we will
adopt it for most of our analyses. Notice that a small fraction of
channeling is exactly what would easily reproduce a clear agreement between DAMA and CoGeNT results also in the standard case of a point--like interactions (see for instance the
results in Ref. \cite{Belli:2011kw}). Therefore to allow for a variation of $f_{\rm ch}$ appears to be
a useful approach, when comparing results from different experiments.

We consider the whole set of DAMA/NaI \cite{dama2008} and DAMA/LIBRA \cite{dama2010} data, which correspond to a cumulative exposure of 1.17 ton$\times$yr.
We analyze the modulation amplitudes $S^{\rm exp}_{mk}$ reported in Fig. 6 in Ref. \cite{dama2010} by using our statistical technique discussed above. The modulation amplitudes of 
Ref. \cite{dama2010} can be considered as a data--reduction of the time-- and energy--
dependent data in 8 energy bins. The actual values of $S^{\rm exp}_{mk}$ of Ref. \cite{dama2008} are valid under the assumption that annual modulation occurs
with phase and period fixed at day 152.5 and 365 days, respectively.
We do not use directly the time--series of the data (which would be a better option for
our statistical technique of studying the null hypothesis) since these are available
only in three energy bins in Refs. \cite{dama2008,dama2010}, while instead the $S^{\rm exp}_{mk}$ are provided in
8 energy bins \cite{dama2010}, therefore supplying more information especially
for light DM. We checked that the results
obtained by using the $S^{\rm exp}_{mk}$ is in full agreement with the results
of Refs. \cite{Belli:2011kw}, where the same type of statistical analysis has been employed directly
on the DAMA data.

In addition, we use the information on the total rate as a constraint, by requiring that the DM contribution $S_{0}$ does not exceed the corresponding experimental value $S^{\rm exp}_{0}$ in the 2--4 keV energy range, measured by DAMA \cite{Nozzoli}. 
Summarizing, for the DAMA datasets our approach requires to determine:
\begin{eqnarray}
 \label{eq:yDAMA}
y &=& -2\ln{\mathcal{L}} ~\equiv ~\chi^2(\epsilon,m_\phi,m_\chi) = \\
 &=& \sum_{k=1}^8\frac{\left(S_{mk} - S_{mk}^{\rm exp}\right)^2}{\sigma_k^2} + 
 \frac{\left(S_{0} - S_{0}^{\rm exp}\right)^2}{\sigma^2} \Theta(S_{0}-S^{\rm exp}_{0}) ,
 \nonumber
\end{eqnarray}
where $\sigma_k$ and $\sigma$ are the experimental errors on $S_{mk}^{\rm exp}$ and 
$S^{\rm exp}_{0}$, respectively, and $\Theta$ denotes the Heaviside
function. The last term in Eq. (\ref{eq:yDAMA}) implements the upper bound on the unmodulated component of the rate $S_0$, by penalizing the likelihood when $S_0$ exceeds $S^{\rm exp}_0$.
For the detector energy--resolution
we use a Gaussian function of width $\sigma_{\rm res}(E)=E(0.448/\sqrt{E}+0.0091)$ \cite{Bernabei:2008yh}.

\subsection{CoGeNT}
\label{sec:cogent}

The CoGeNT experiment is made by Ge detectors with very low threshold (about $0.4$ keVee, where keVee denotes keV electron--equivalent energy). 
Thanks to this property, CoGeNT has the capability of being very sensitive to DM candidates with $m_\chi\lesssim 10$ GeV, although large background contamination may be present
at these low--energies.

In 2010, the CoGeNT Collaboration reported the detection of an excess not identifiable with a known background, and potentially compatible with a DM interpretation~\cite{cogentrate}.
More recently, the temporal evolution of the measured rate in different energy bins for data taken between December 4, 2009 and March 6, 2011 (442 live-days) has been 
presented \cite{cogentmod} . At low energies, CoGeNT data favour the presence of an annual modulation which can be fitted by a WIMP signal \cite{cogentmod}.

One could perform an analysis similar to the one realized for DAMA by deriving
the modulation amplitudes $S^{\rm exp}_{mk}$ from the data presented by the
COGeNT Collaboration in Fig.~4 of Ref. \cite{cogentmod}. To this aim we fix the
period of modulation at 365 days and the phase at June 2nd (day 152.5), as for the
DAMA dataset.
We obtain $S^{\rm exp}_{m1}=0.12 \pm 0.08 $ counts/day in the 0.5--0.9 keV energy bin and $S^{\rm exp}_{m2}= 0.26 \pm 0.17$ counts/day in the 0.9--3.0 keV energy bin.
However, as mentioned above, for our statistical technique (test of the null hypothesis) acting directly on the time–series of the data is more appropriate, and so 
for CoGeNT we follow this path.
In addition to the modulation amplitude, we treat the total rate measure by CoGeNT as a constraint and, similar to the case of the analysis of the DAMA data, we define:

\begin{eqnarray}
 \label{eq:ycogent}
&& y = -2\ln{\mathcal{L}} ~\equiv~ \chi^2(\epsilon,m_\phi,m_\chi) = \\
&& \sum_{k=1}^{16}\frac{\left(\tilde S_{m1,k} - \tilde S_{m1,k}^{\rm exp}\right)^2}{\sigma_k^2} +
  \sum_{k=1}^{16}\frac{\left(\tilde S_{m2,k} - \tilde S_{m2,k}^{\rm exp}\right)^2}{\sigma_k^2} + \nonumber \\
&& \sum_{j=1}^{31}\frac{\left(S_{0j} - S_{0j}^{\rm exp}\right)^2}{\sigma_j^2} \Theta(S_{0j}-S^{\rm exp}_{0j}) ,
 \nonumber
\end{eqnarray}
where $\tilde S_{mk}=1/\Delta t_k \int_{\Delta t_k} dt\,S_{mk} \cos\left[2\pi(t-\phi)/\tau\right] $, with $\Delta t_k$ being the temporal bin of experimental data, and $\tilde S^{exp}_{mk}=R^{exp}_{mk}-<R^{exp}_m>$ with $R^{exp}_{mk}$ being the total rate shown in Fig. 4 of Ref. \cite{cogentmod} and $<R^{exp}_m>$ being the rate $R^{exp}_{mk}$ averaged over a cycle (1 year). We compute the total rate in the $0.9-3.0$ keV energy-bin by simply subtracting the rate in $0.5-0.9$ keV bin to the rate in the $0.5-3.0$ keV bin, and with a Gaussian propagation of errors. 
$S^{\rm exp}_{0j}$ and $\sigma_j$ 
denote the counts and their corresponding errors given in Ref. \cite{cogentmod} 
(31 bins in the energy interval $0.4 - 2$ keVee), with
the L--shell peaks removed, but without any further background removal.

Notice that with the procedure of Eq. (\ref{eq:ycogent}) we do not require that the whole
total (unmodulated) spectrum of CoGeNT is due to DM scattering: we just require that the the
spectrum is not exceeded by our theoretical model. The unmodulated spectrum $S^{\rm exp}_{0j}$ acts therefore as a bound, leaving room in it for an unknown background component.

The total fiducial mass of the CoGeNT experiment is 330 g, the energy resolution is described by a Gaussian function with the form of the width $\sigma_{\rm res}$ taken from \cite{Aalseth:2008rx}, and the quenching factor follows from the relation $E=0.2\,E_{\rm R}^{1.12}$ below 10 keV~\cite{Barbeau:2007qi}. 

\subsection{CRESST}
The CRESST experimental setup~\cite{Jochum:2011zz} at the Gran Sasso Laboratories includes 300 g of scintillating CaWO$_4$ target crystals. 
The particle interaction is detected through phonons by the phase transition thermometer and through scintillation light by a separate cryogenic detector.
Results from 730 kg days of data--taking have been recently presented by the CRESST collaboration~\cite{Angloher:2011uu}.
Sixty-seven events are found in the WIMP acceptance region, and background contributions from leakage of e/$\gamma$-events, neutrons,
$\alpha$-particles, and recoiling nuclei in $\alpha$-decays are not sufficient to explain all the observed events. A likelihood--ratio test
rejects the background--only hypothesis at a significance larger than 4-$\sigma$~\cite{Angloher:2011uu}.

To perform our analysis, we compute DM signal--events in each of the 8 CRESST detector modules. We consider acceptance regions and number of observed events, as provided in Table 1 of Ref.~\cite{Angloher:2011uu}, and background events are derived according to estimates in Sec.~4 of Ref.~\cite{Angloher:2011uu}. Performing a likelihood--ratio test, we obtain an evidence for the best--fit DM signal over a background--only scenario at 4.1-$\sigma$ C.L., thus in good agreement with the result of the Collaboration.

However, we do not have at disposal all the information needed to perform a full analysis, so our derived contours and allowed regions must then be considered as indicative.
In particular we use the published exposure of 730 kg$\times$days assuming an equal contribution from each module (which thus has an exposure of 730/8 kg$\times$days) and consider a constant efficiency.
Moreover, in order to be able to properly discriminate among DM models inducing different recoil spectra, we would need the energies of events in each detector, rather than the total number in the whole acceptance region. This is the reason why our allowed regions for
the standard point--like case do not overlap with the regions presented in Ref.~\cite{Angloher:2011uu} for large dark matter masses. On the other hand, at masses below 50 GeV (which is the main focus of this paper), the agreement becomes very good.

The likelihood function $y$ is used as described above to determine the allowed regions in the parameter space of our model.

\subsection{CDMS}

The CDMS experiment, located at the Soudan Underground Laboratory in Minnesota (like CoGeNT), has operated cryogenic semi--conductor detectors to simultaneously measure phonons and charge in order to reject most of the dominant radioactive background and disentangle a DM signal. Only the Ge detector were completely functioning and fully exploited to set constraints on DM properties.

In this paper, we consider the `standard' 2009 CDMS-II results~\cite{Ahmed:2009zw}. They are based on Ge data taken between July 2007 and September 2008, applying conservative nuclear recoil selection cuts and assuming a 10 keV energy threshold. The total exposure is 612 kg$\times$ days and efficiency has been taken from Fig.~5 in Ref. \cite{Ahmed:2010hw} (black curve) with quenching factor $q\simeq1$.\footnote{For the cases of interests (low DM masses), a similar analysis can be performed by exploiting combined data from CDMS and EDELWEISS experiments (see Fig. 1 in Ref. \cite{CDMS:2011gh}), obtaining basically the same results.}
Two signal events were seen in the 10--100 keV energy window~\cite{Ahmed:2009zw} against an expected background of $0.9\pm 0.2$ (which are the numbers we use to derive constraints).

\subsection{XENON100}
The XENON100 experiment searches for DM scatterings on a target of purified liquid Xe by measuring scintillation and ionization signals.
We consider results presented in Ref. \cite{xenon100}, corresponding to an exposure of 100.9 days in a 48 kg fiducial volume. After the implementation of all the cuts
three events have been reported in the DM signal region with an expected background of $1.8 \pm 0.6$ events.

The XENON collaboration selected the energy window for the WIMP search region to be between 4--30 photoelectrons (in terms of prompt scintillation light in the liquid), corresponding to recoil energy of 8.4 -- 44.6 keVnr (based on the $\mathcal{L}_{\rm eff}$ parametrization in Ref.\cite{xenon100}). However, recoils at lower energy can contribute as well, especially 
close to threshold, due to the Poissonian tail.
Both the statistics and the quenching of few keVnr nuclear recoils in liquid Xenon are not completely understood (for recent discussions, see {\rm e.g.}, Ref. \cite{collar/xenon} and reference therein). For definiteness, we consider a Poissonian distribution of photoelectrons, and include a single--photoelectron resolution of 0.5.
The $\mathcal{L}_{\rm eff}$ function is a very crucial ingredient at this low level of
photoelectrons and for light DM. We adopt two different approaches, in order to bracket  a possible (but definitely not exhaustive) uncertainty on the derived bounds from XENON100:
i) we adopt the nominal central value of $\mathcal{L}_{\rm eff}$
shown in Fig.1 of Ref. \cite{xenon100}, which heavily relies on linear extrapolation
below 3 keVnr; ii) more conservatively, we increase the photomultiplier
threshold to 8 photoelectrons, in order to determine a situation which is nearly independent
on the knowledge of $\mathcal{L}_{\rm eff}$ below 3 keVnr. The value of 8 photoelectrons has been chosen to this purpose (namely, it is the lowest value satisfying such requirement). Notice that these two
recipes do not exhaust the possibilities of alternative assumptions that can be done
to determine the XENON 100 response to light DM. For more discussion and additional considerations, see {\em e.g.} Refs. \cite{dama/xenon,collar/xenon}. Due to the large uncertainties inherent in the
derivation of bounds from XENON100 for light DM, it appears to be still preliminary to
assume those bounds as strictly firm. We nevertheless show them, but without enforcing them in our discussion. In any case, conservatively, we consider the 8 PE bound as more appropriate since it is less dependent on the $\mathcal{L}_{\rm eff}$ extrapolation.

Finally, to compute the expected signal we follow Eq. (13--16) in Ref. \cite{Aprile:2011hx}.
In both the cases of CDMS and XENON, we derive upper bounds as discussed at the beginning
of Section \ref{sec:analysis}.

\section{Results}

\begin{figure*}[t]
\includegraphics[width=0.49\textwidth]{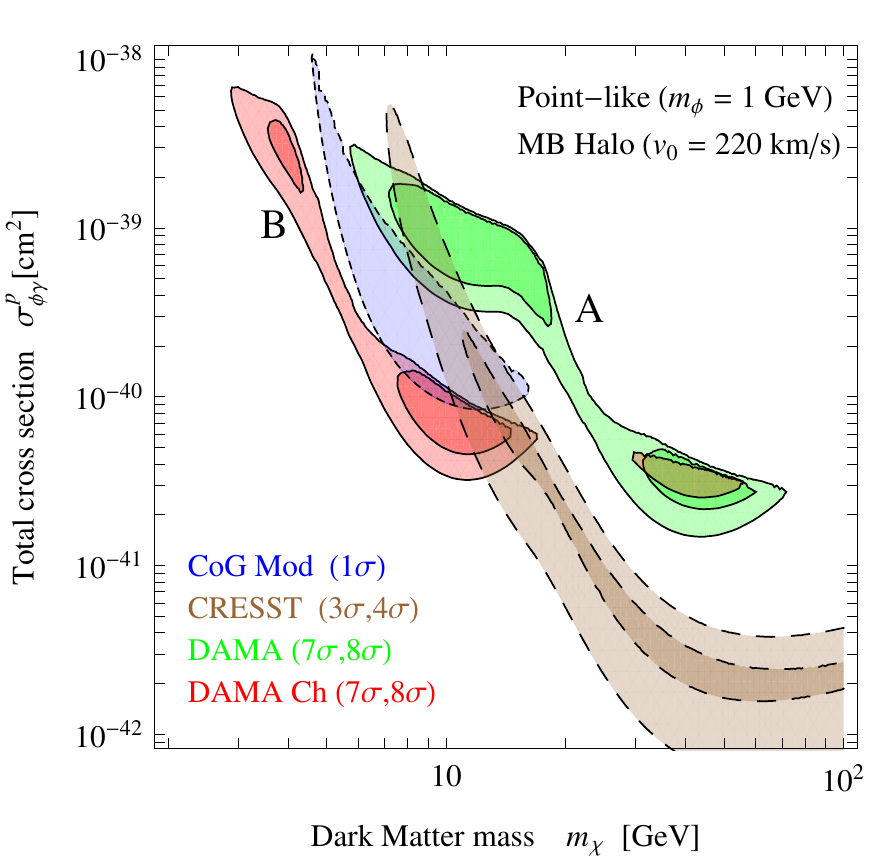}
\includegraphics[width=0.49\textwidth]{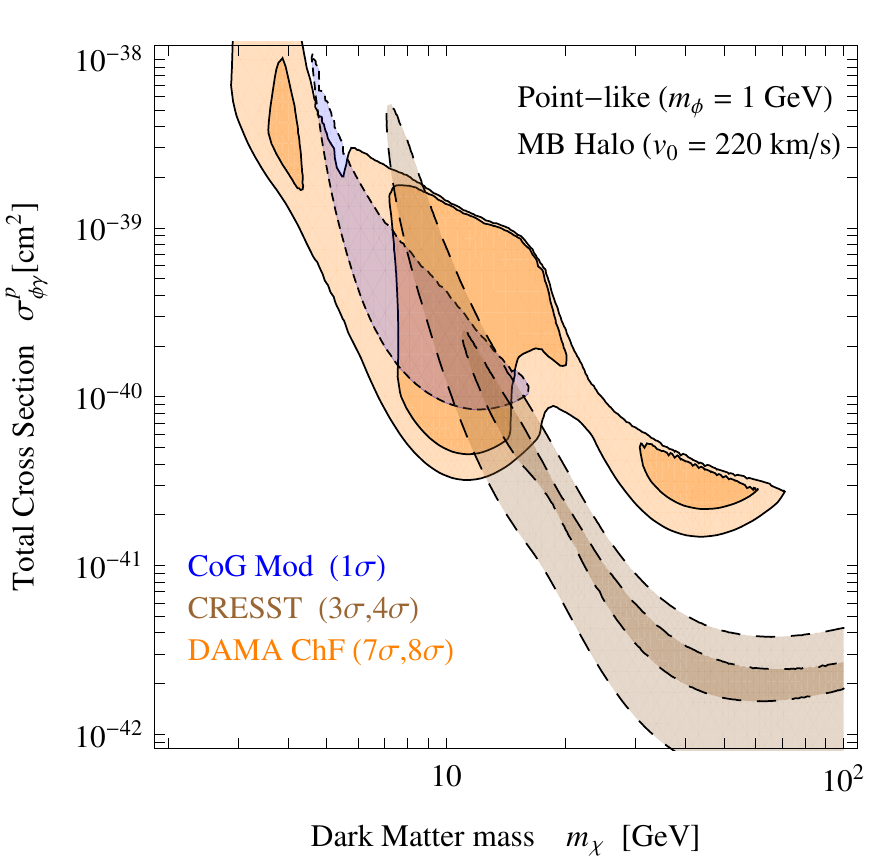}
\caption{Point--like scattering cross sections on proton, as a function
of the dark matter mass. The galactic halo has been assumed in the form of an isothermal
sphere with velocity dispersion $v_0=220$ km s$^{-1}$ and local density
$\rho_0 = 0.3$ GeV cm$^{-3}$. 
{\sc Left panel:} The solid green contours A, denote the
regions compatible with the DAMA annual modulation effect \cite{dama2008,dama2010}, in 
absence of channeling. The solid red contours B, refer to the regions compatible with the
DAMA annual modulation effect, when the channeling effect is considered at its maximal
value. 
The dotted blue contour refers to the region derived from the CoGeNT annual modulation 
effect \cite{cogentmod}, when the bound from the unmodulated CoGeNT data is included.
 The dashed brown contours denote the regions compatible with the CRESST excess \cite{Angloher:2011uu}.
For all the data sets, the contours refer to regions where the absence of modulation
can be excluded with a C.L. of: $7\sigma$ (outer region), $8\sigma$ (inner region) for DAMA, 
$1\sigma$ for CoGeNT and  $3\sigma$ (outer region), $4\sigma$ (inner region)
for CRESST.
{\sc Right panel:} The same as in the left panel, with the following difference: the solid
orange contour refers to the DAMA annual modulation data, when the fraction of channeling is varied in its allowed interval \cite{Bernabei:2007hw}. 
Again, the contour refers to the region where absence of modulation
can be excluded with a C.L. $7\sigma,8\sigma$.}
\label{fig:B}
\end{figure*}

\begin{figure*}[t]
\includegraphics[width=0.49\textwidth]{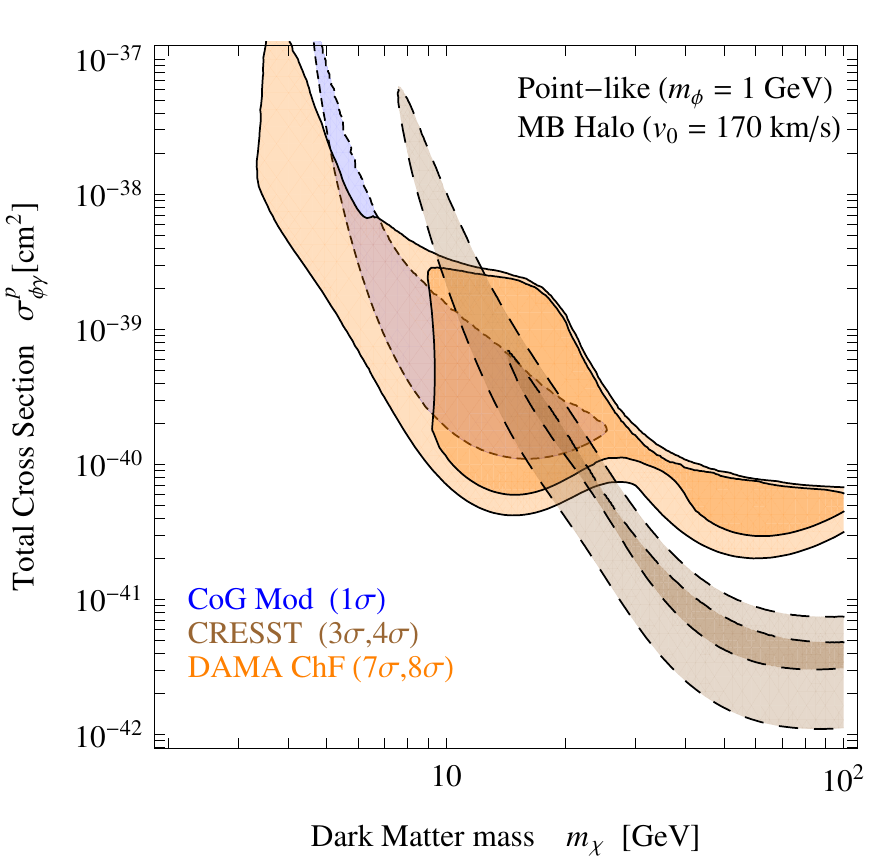}
\includegraphics[width=0.49\textwidth]{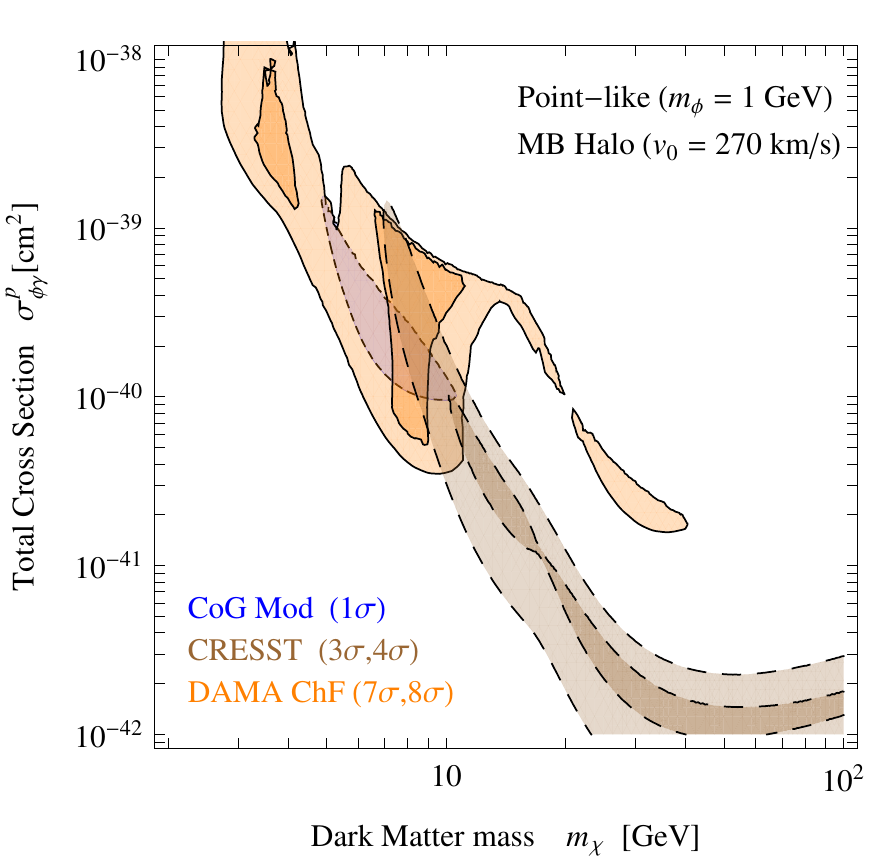}
\caption{Point--like scattering cross sections on proton, as a function
of the dark matter mass. The galactic halo has been assumed in the form of an isothermal
sphere with velocity dispersion $v_0=170$ km s$^{-1}$ and local density
$\rho_0 = 0.18$ GeV cm$^{-3}$ (left panel);  
$v_0=270$ km s$^{-1}$ and local density
$\rho_0 = 0.45$ GeV cm$^{-3}$ (right panel). Notations are the same as in the left panel
of Fig. \ref{fig:B}.}
\label{fig:B-more}
\end{figure*}

\begin{figure*}[t]
\includegraphics[width=0.49\textwidth]{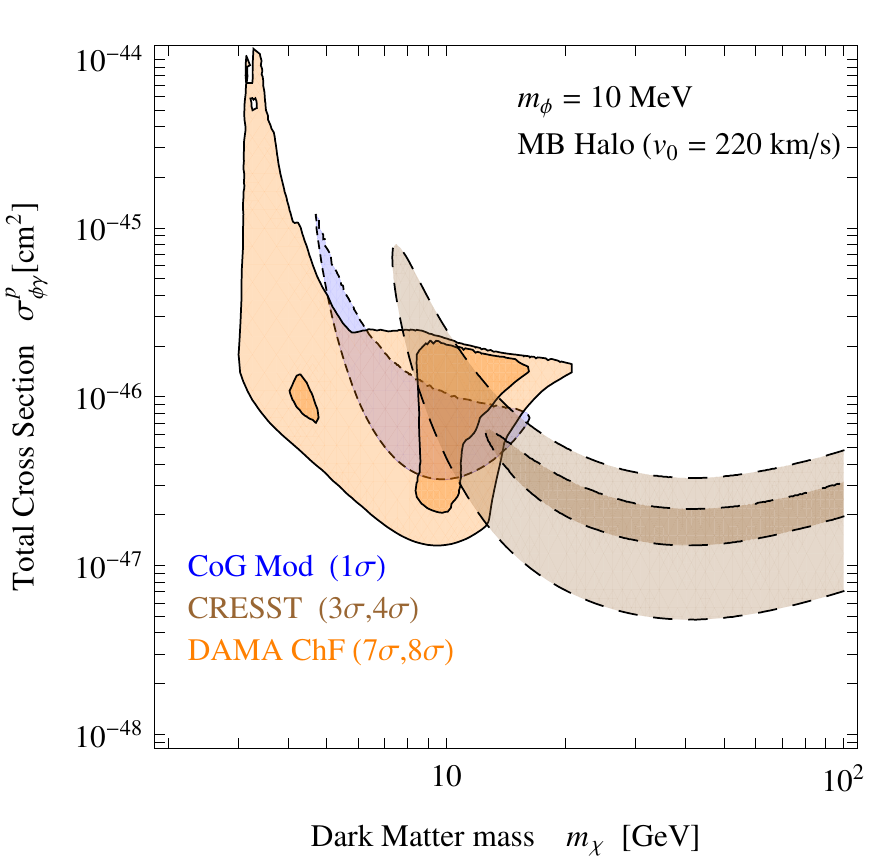}
\includegraphics[width=0.49\textwidth]{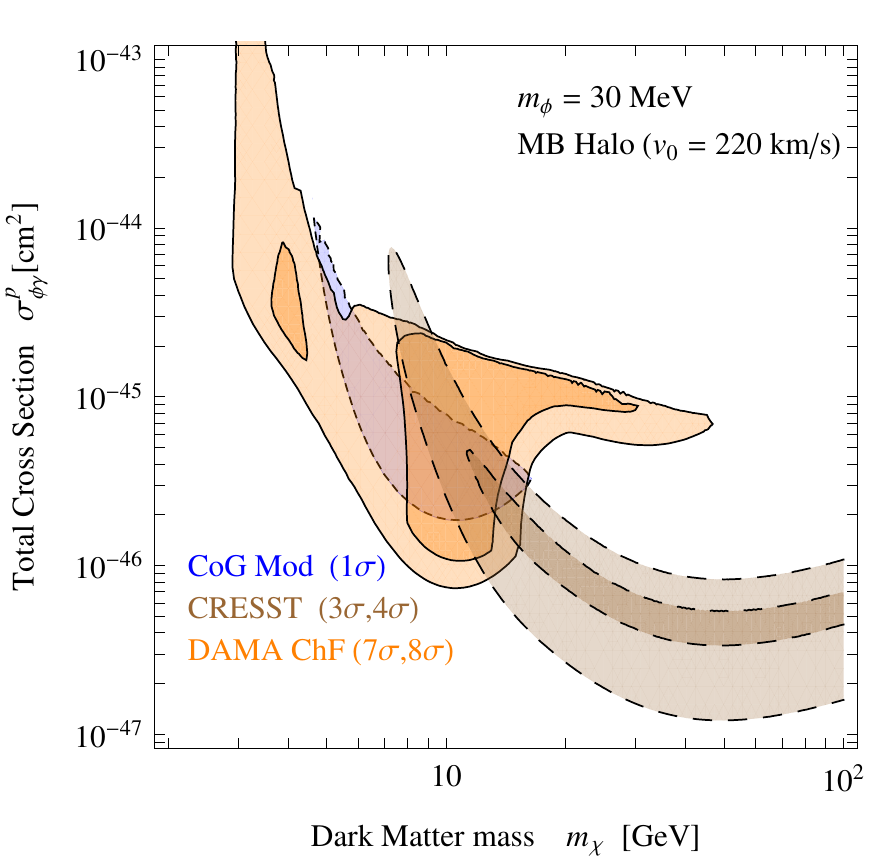}
\caption{Scattering cross sections on proton, as a function
of the dark matter mass, for a long--range mediator of mass $m_\phi = 10$ MeV
(left panel) and 30 MeV (right panel). 
The galactic halo has been assumed in the form of an isothermal
sphere with velocity dispersion $v_0=220$ km s$^{-1}$ and local density
$\rho_0 = 0.3$ GeV cm$^{-3}$. 
Notations are the same as in the left panel
of Fig. \ref{fig:B}.
}
\label{fig:C}
\end{figure*}

\begin{figure*}[t]
\includegraphics[width=0.49\textwidth]{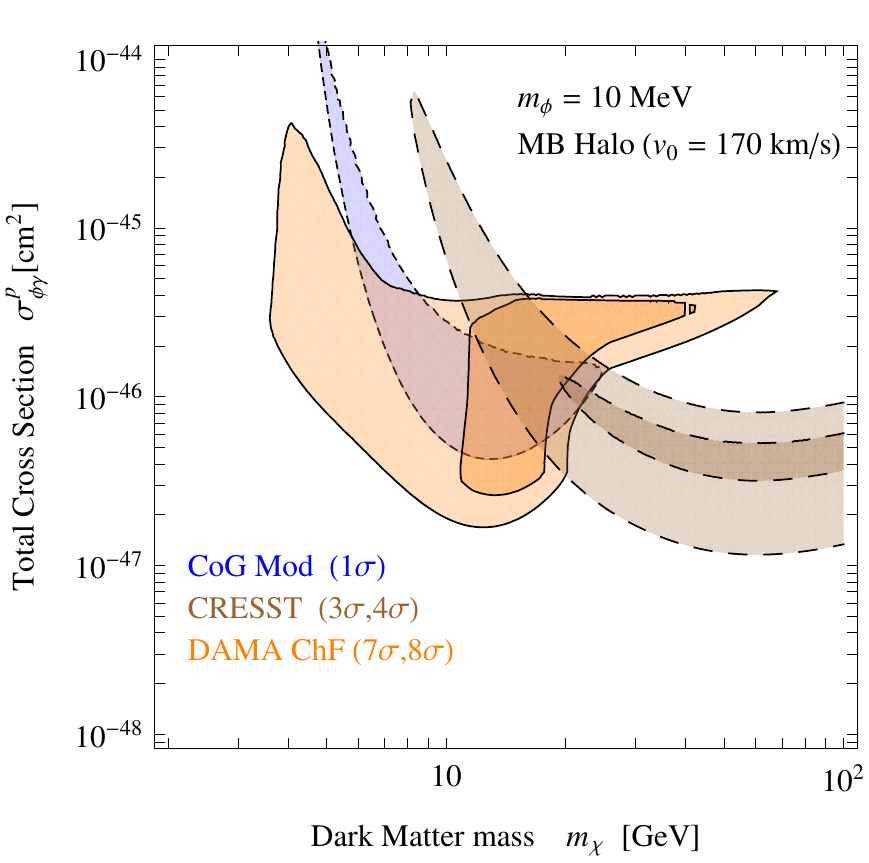}
\includegraphics[width=0.49\textwidth]{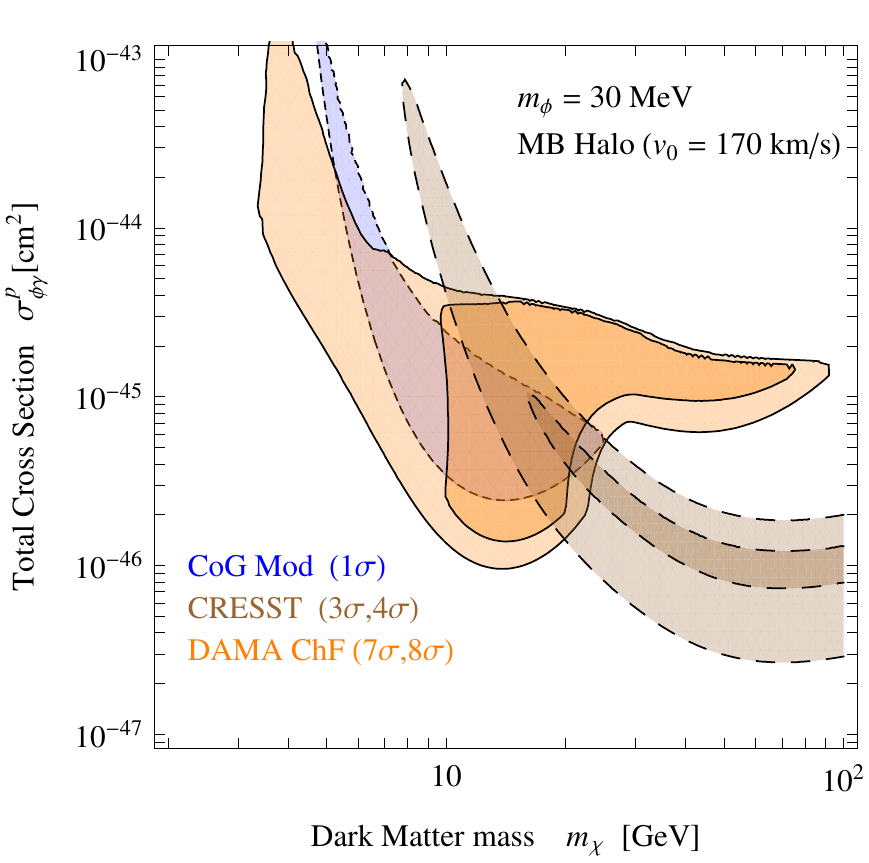}
\caption{The same as in Fig. \ref{fig:C}, for an isothermal
sphere with velocity dispersion $v_0=170$ km s$^{-1}$ and local density
$\rho_0 = 0.18$ GeV cm$^{-3}$. 
}
\label{fig:D}
\end{figure*}

\begin{figure*}[t]
\includegraphics[width=0.49\textwidth]{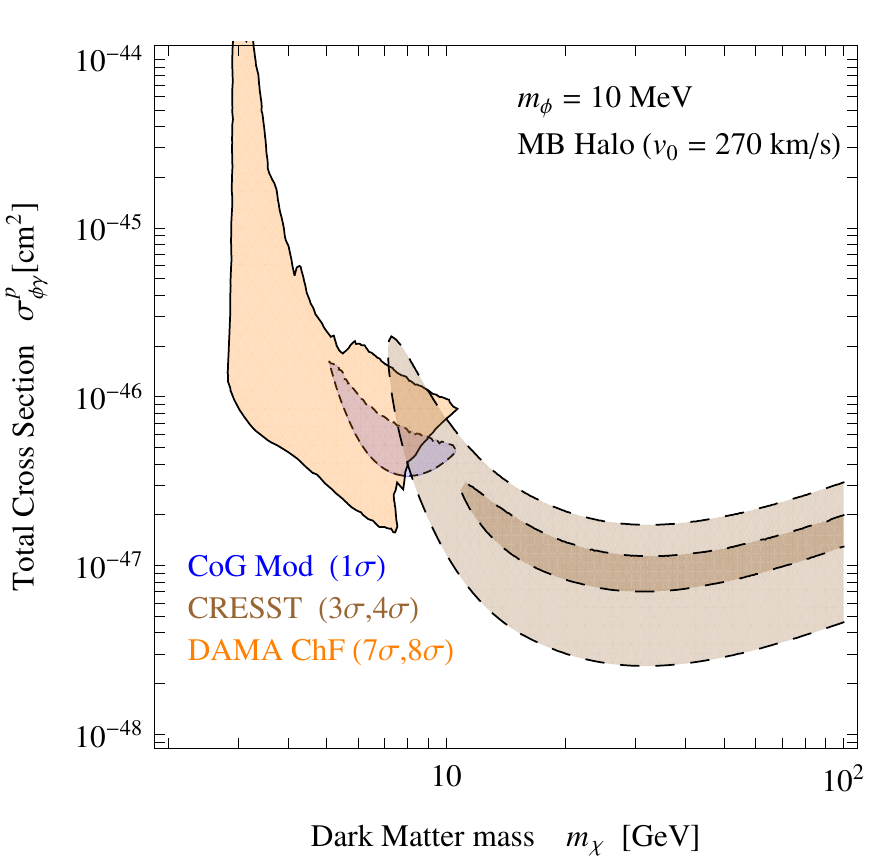}
\includegraphics[width=0.49\textwidth]{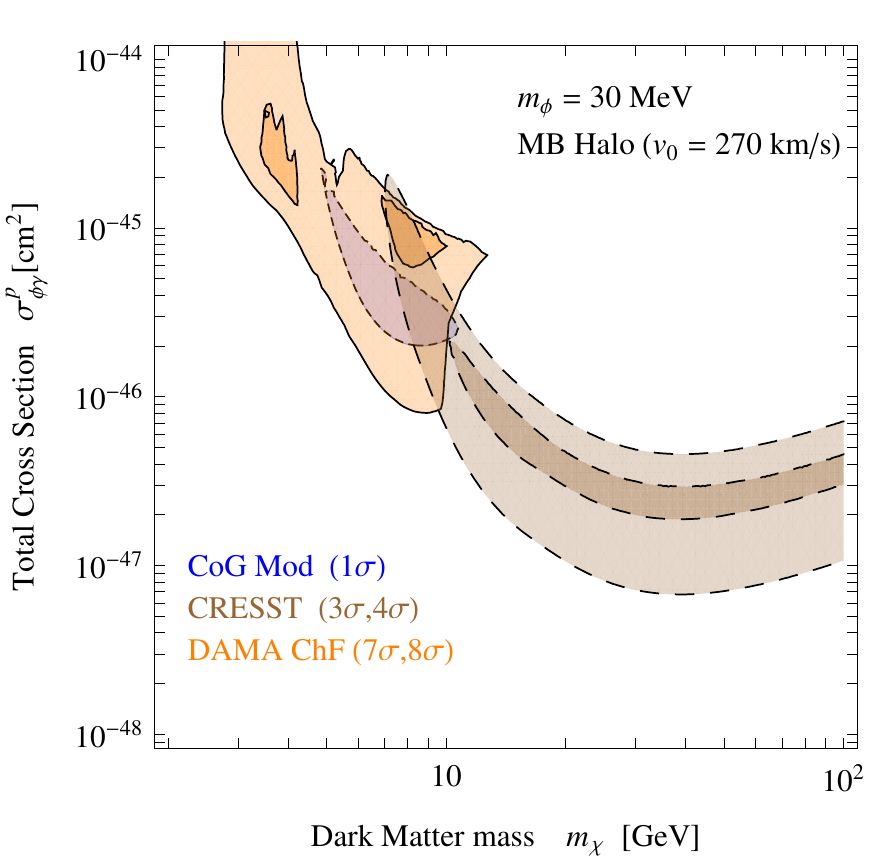}
\caption{The same as in Fig. \ref{fig:C}, for an isothermal
sphere with velocity dispersion $v_0=270$ km s$^{-1}$ and local density
$\rho_0 = 0.45$ GeV cm$^{-3}$. 
}
\label{fig:E}
\end{figure*}

\begin{figure}[t]
\includegraphics[width=0.49\textwidth]{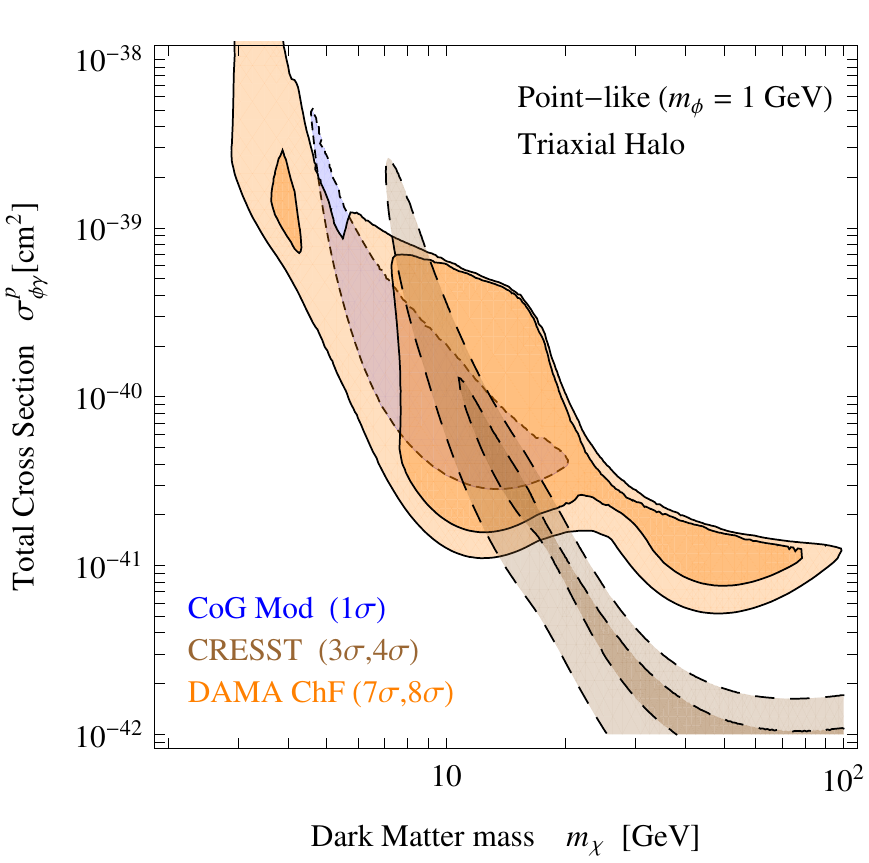}
\caption{The same as in Fig. \ref{fig:C}, except that the interaction is point--like, 
and for
a triaxial galactic halo
with the Earth located on the major axis \cite{Evans:2000gr}, with velocity dispersion 
$v_0=220$ km s$^{-1}$ and local density
$\rho_0 = 0.84$ GeV cm$^{-3}$ \cite{Belli:2002yt}.}
\label{fig:F}
\end{figure}


Let us now move to the discussion of our results in terms of long--range interactions
of DM on protons. We show in this Section the effect of moving from the point--like case (parameterized, for convenience, in terms of a massive mediator $\phi$, for which we use a reference value for its mass of 1 GeV, larger values reproducing
the same results) to the very long range case (which refers to the extreme case of $m_\phi = 0$), going through the intermediate case of relatively light $\phi$. To this aim, we sit
where the sensitivity to $m_\phi$ is largest, which is for values ranging from several
MeV's to few tens of MeV's. This fact can be seen, in an illustrative case, in Fig. \ref{fig:A}, where we show, in the plane $\epsilon$ vs. $m_\phi$,
the iso--contours of constant rate (chosen as 1 cpd/kg) on a Na target, for a 10 GeV
DM--particle scattering and for various values for the energy threshold of a Na detector (for definiteness, we set here $Z'=1$ and $\alpha_{\rm dark}=\alpha_{SM}$).
 As discussed in Section \ref{sec:signals},
for values of the mediator mass smaller that a few MeV (the actual
value depending on the recoiling nucleus and detected energies) the differential cross section and the corresponding experimental rate become proportional to $\epsilon^2/E_R^2$, typical of a very long--range interaction, and the dependence of the iso--contours
of constant rate becomes effectively insensitive to $m_\phi$. 
Fig. \ref{fig:A} shows that this long--range limit
is reached, for typical nuclei and recoil energies in the range of interest
of actual experiments, for $m_\phi \lsim 1$ MeV. On the contrary,
for values of $m_\phi$ larger that several tens of MeV,
the cross section reaches the point--like limit: in this
case the detection rate is proportional to $\epsilon^2/m_\phi^{4}$ and the iso--contours
of constant rate follow the linear behaviour (in log--log scale) shown in the figure.
The transition regime is then obtained for $m_\phi$ for values ranging from few MeV to few tens of MeV. For definiteness, in our analyses we will adopt the values of $m_\phi = 10$ MeV and 30 MeV. 

Since the response of direct detection experiments is quite sensitive to the
DM distribution in the galactic halo \cite{Belli:2002yt}, especially in velocity space, 
we perform our analysis by considering two kinds of variations in this respect. 

First, we analyze the direct detection datasets by using a standard isothermal model, which basically implies a truncated Maxwell--Boltzmann velocity distribution function, but we take into account uncertainties
on the velocity dispersion $v_0$, as discussed in Ref. \cite{Belli:2002yt}. 
We recall that in the case of an isothermal model, the DM velocity dispersion is directly linked to the
asymptotic value of the rotational velocity supported by the DM halo: uncertainties in the
velocity dispersion $v_0$ are therefore representative of the uncertainties in the
local rotational velocity \cite{Belli:2002yt}. Following Ref. \cite{Belli:2002yt}
we will use the three values $v_0 = 170, 220, 270$ km s$^{-1}$, which bracket the
uncertainty in the local rotational velocity.
 Let us notice that also the value of the local DM density $\rho_0$ is correlated to the adopted value of $v_0$, as discussed {\em e.g.} in Ref. \cite{Belli:2002yt}. This case corresponds to the model denoted
as A0 in Ref. \cite{Belli:2002yt}, and we adopt the case of minimal halo, which implies
lower values of local DM density (since a fraction of the galactic potential is supported
by the disk/bulge). In turn, this implies the adoption of $\rho_0 = 0.18, 0.30, 0.45$ GeV 
cm$^{-3}$ for $v_0 = 170, 220, 270$ km s$^{-1}$, respectively \cite{Belli:2002yt}.

Second, we consider a different form for
the galactic halo, and in order to somehow emphasize the difference with the isothermal
sphere, we adopt a triaxial halo model \cite{Evans:2000gr}, with anisotropic velocity dispersions. 
We adopt the model denoted by D2 in Ref. \cite{Belli:2002yt}, which corresponds to the case
when the Earth is located on the major axis of the potential ellipsoids, with
$v_0 = 220$ km s$^{-1}$ and maximal halo, which implies $\rho_0 = 0.84$ GeV 
cm$^{-3}$ \cite{Belli:2002yt}.

\subsection{Point--like scenarios}

\begin{figure*}[t]
\includegraphics[width=0.49\textwidth]{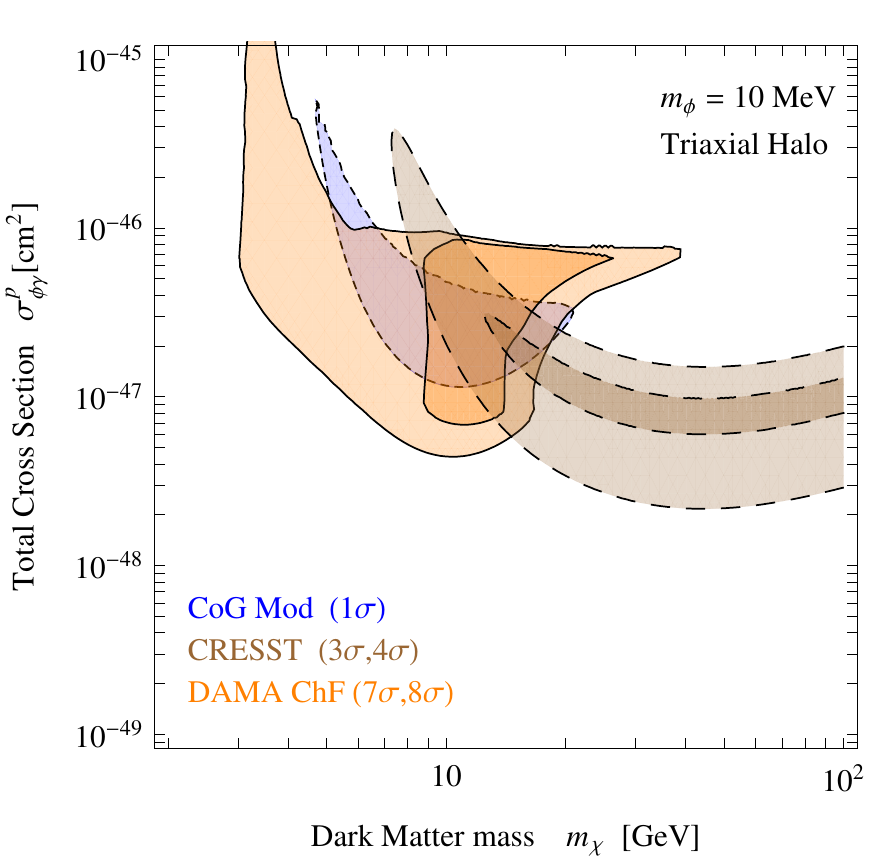}
\includegraphics[width=0.49\textwidth]{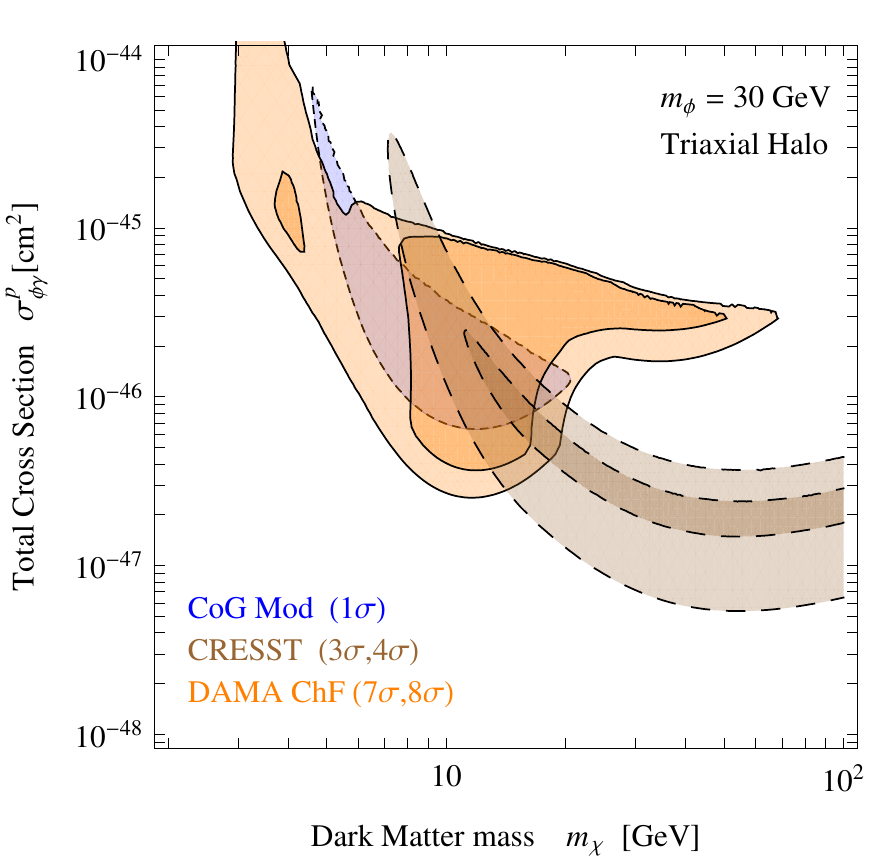}
\caption{The same as in Fig. \ref{fig:C}, for a triaxial galactic halo
with the Earth located on the major axis \cite{Evans:2000gr}, with velocity dispersion 
$v_0=220$ km s$^{-1}$ and local density
$\rho_0 = 0.84$ GeV cm$^{-3}$ \cite{Belli:2002yt}.  
}
\label{fig:G}
\end{figure*}


We start with the point--like scenario. Fig. \ref{fig:B} shows the scattering cross sections on proton $\sigmap$ as a function
of the dark matter mass $m_\chi$. The galactic halo has been assumed in the form of an isothermal
sphere with velocity dispersion $v_0=220$ km s$^{-1}$ and local density
$\rho_0 = 0.3$ GeV cm$^{-3}$. In the left panel we show the allowed regions compatible
with the annual modulation effects in DAMA, the CoGeNT excess, as well as  the region which
turns out to be compatible with the CRESST excess, when interpreted as a DM signal.
Specifically, the solid green contours A, denote the regions compatible with the DAMA annual modulation effect \cite{dama2008,dama2010}, in 
absence of channeling. The solid red contours B, refer to the regions compatible with the DAMA annual modulation effect, when the channeling effect is considered at its maximal value.\footnote{DAMA is a multi-target detector and allowed regions at large DM mass correspond to scattering on I, while at small DM mass regions correspond to scattering on Na. Note that the region at DM mass around 10 GeV is given by scattering on Na targets in the no-channeling case, and by scattering on I targets in the maximal channeling case.}
The dotted blue contour refers to the region derived from the CoGeNT annual modulation effect \cite{cogentmod}, when the bound from the unmodulated CoGeNT data is included. The dashed brown contours denote the regions compatible with the CRESST excess \cite{Angloher:2011uu}. 
For all the data sets, the contours refer to regions where the absence of modulation
can be excluded with a C.L. of: $7\sigma$ (outer region), $8\sigma$ (inner region) for DAMA, 
$1\sigma$ for CoGeNT and $3\sigma$ (outer region), $4\sigma$ (inner region)
for CRESST.  The right panel, instead, shows a further analyses for the DAMA data: the solid orange contours refer to the results obtained by varying the channeling fraction $f_{\rm ch}$ in its allowed range, as discussed in Section \ref{sec:DAMA}. We can therefore see the extent of the DAMA allowed region
when $f_{\rm ch}$ is marginalized over. We will adopt this procedure of treating
the channeling effect in DAMA
in the remainder of the paper. 

From Fig. \ref{fig:B} we notice that, in the case of a point--like cross section, the DAMA and CoGeNT regions both point toward a DM with a mass in the 10 GeV ballpark 
(more specifically, from about 5 up to about 20 GeV)
and cross
sections from a few $\times 10^{-41}$ cm$^2$ to $10^{-38}$ cm$^2$ (with our choice of galactic halo model and for our situation of scattering on protons only), a result which confirms similar analyses \cite{Belli:2011kw,Foot:2011pi,Schwetz:2011xm,Farina:2011pw,McCabe:2011sr,Fox:2011px}. The DAMA and CoGeNT regions
are largely overlapped. This is especially clear in the right panel of Fig. \ref{fig:B}, where
we marginalize over the channeling fraction: even a small amount of channeling is enough
to make DAMA and CoGeNT regions perfectly overlap. 

It is interesting to note that, when channeling is introduced, there is an allowed region for DM mass of 4--5 GeV, which corresponds to the signal given by the scattering of DM particles on the fraction of Na targets undergoing channeling. This region is very seldom considered in analyses of DAMA data, and the reason resides in the statistical technique adopted. Indeed, the statistical evidence associated to this region is very high when compared to the absence of signal (up to 8--$\sigma$), but its $\chi^2$ is substantially larger than the best--fit DM model (which is for DM mass around 10 GeV). Therefore, when regions are drawn assuming that a DM signal is present and including models that falls within a certain C.L. from the best--fit case (as often done in the literature, except in the analysis performed by the
DAMA Collaboration itself, see e.g. Refs. \cite{Bernabei:2003za,Bernabei:2007hw}
and Refs. \cite{Belli:2002yt,Belli:2011kw}) such region disappears (unless allowing extremely large C.L. with respect to the minimum $\chi^2$).
This is also the case in our Fig.~13, where we show ``preferred'' regions, i.e., domains which include models having a $\chi^2$ within a given C.L. with respect to the best--fit $\chi^2$.
An allowed region for DM mass of 4--5 GeV is particularly interesting, especially in light of the fact that it is compatible with both DAMA and CoGeNT data, and can easily satisfy constraints from other experiments, such CDMS and XENON100 (independently on the method employed to derive the latter).

The statistical significance for the presence of a signal due to DM scattering 
is reported in Table \ref{tab1}, which confirms that for  
DAMA the interpretation in terms of a DM signal is highly favored 
($9.6\sigma$ for the case under consideration, i.e. isothermal sphere with
central value of $v_0$). In the case of
CoGeNT the effect is at the level of $1.8\sigma$, a fact that simply reflects the current lower statistics of the
CoGeNT data sample.

Let us stress that we are not considering here a number of sources of uncertainties of experimental origin which can be relevant in the analysis, like for instance the uncertainties on the quenching factors in Na, I and Ge \cite{Belli:2011kw}. Those effects would somehow
enlarge both the DAMA and CoGeNT (and CRESST, too) regions, allowing a wider range of
DM masses.

Fig. \ref{fig:B} also shows, as brown contours, the preferred region derived from the CRESST excess. We remind that
some assumptions on unknown features of the data were needed. In particular, without energy spectra of recoil events for each module, the discrimination between a large--mass WIMP (nearly flat spectrum) and a small--mass WIMP (exponential spectrum) becomes difficult. The total spectrum shown in Fig.~11 of Ref. \cite{Angloher:2011uu} points toward the latter rather than the former case. Therefore, at small masses, we do not expect CRESST regions to be drastically modified by an analysis with the full set of information. Indeed, our preferred region perfectly overlaps with the one found in \cite{Angloher:2011uu} for DM mass below 50 GeV. The picture is different at large masses, where our CRESST regions 
are affected by our imperfect knowledge on relevant detection information.

In the current situation, the analysis shows that the contours perfectly overlap with the DAMA 
and CoGeNT regions, and it is very intriguing that all three ``positive'' experimental results point to the same sector of DM parameter space.

The effect induced by the variation in the DM dispersion velocity is shown in the
two panels of Fig. \ref{fig:B-more}. In the case of $v_0=170$ km s$^{-1}$, regions are
not significantly modified as compared to the case of $v_0=220$ km s$^{-1}$ (except for the
overall normalization, a fact which reflects the different values of local DM density
in the two cases). Table \ref{tab1} shows that also the statistical significance is similar
to the $v_0=220$ km s$^{-1}$ case. Larger dispersion velocities, instead, are more favorable
for lighter DM masses. The right panel of Fig. \ref{fig:B-more} shows in fact that the allowed regions
are shifted toward lower masses for $v_0=270$ km s$^{-1}$. Regions still overlap,
although they shrink to more defined ranges, both in DM mass and cross section. In this
case, the DM mass cannot extend much above 10 GeV. From Table \ref{tab1} we see that the statistical significance is reduced by about $1\sigma$ for DAMA. Clearly
this level of reduction is not sizable enough to allow exclusion of the $v_0 = 270$ km s$^{-1}$ case.

\subsection{Long--range forces scenarios}

\begin{figure}[t]
\includegraphics[width=0.49\textwidth]{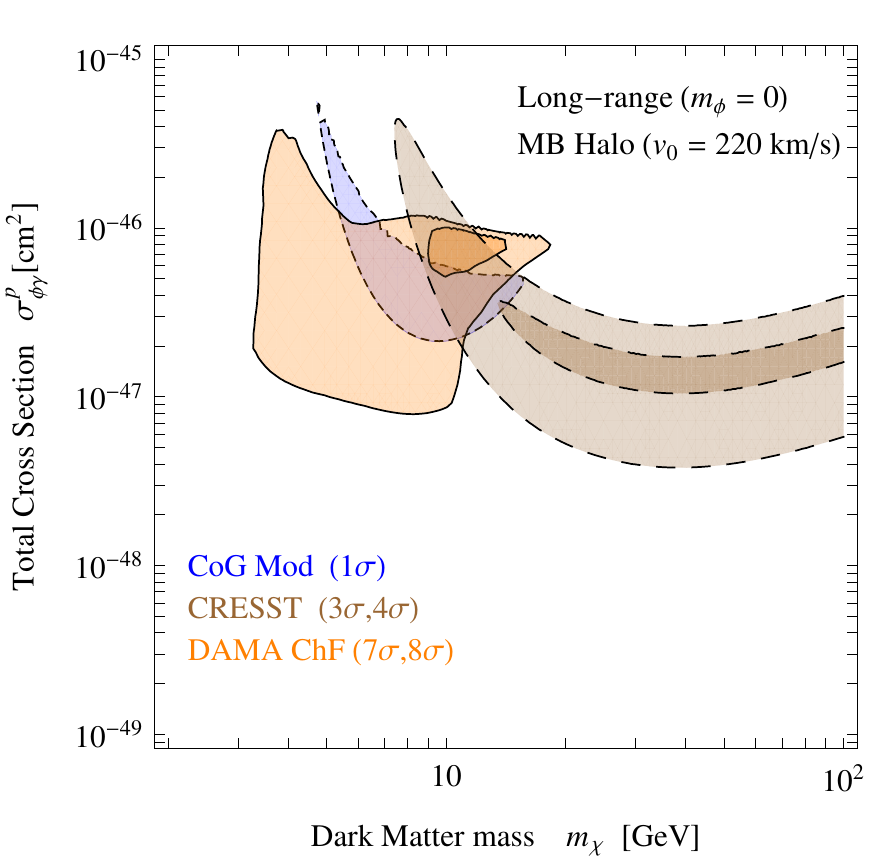}
\caption{The same as in Fig. \ref{fig:C}, except that the interaction is
long--range ($m_\phi = 0$ MeV), for the isothermal sphere and for
a dispersion velocity $v_0=220$ km s$^{-1}$ and local density
$\rho_0 = 0.3$ GeV cm$^{-3}$. 
}
\label{fig:Ha}
\end{figure}

\begin{figure*}[t]
\includegraphics[width=0.49\textwidth]{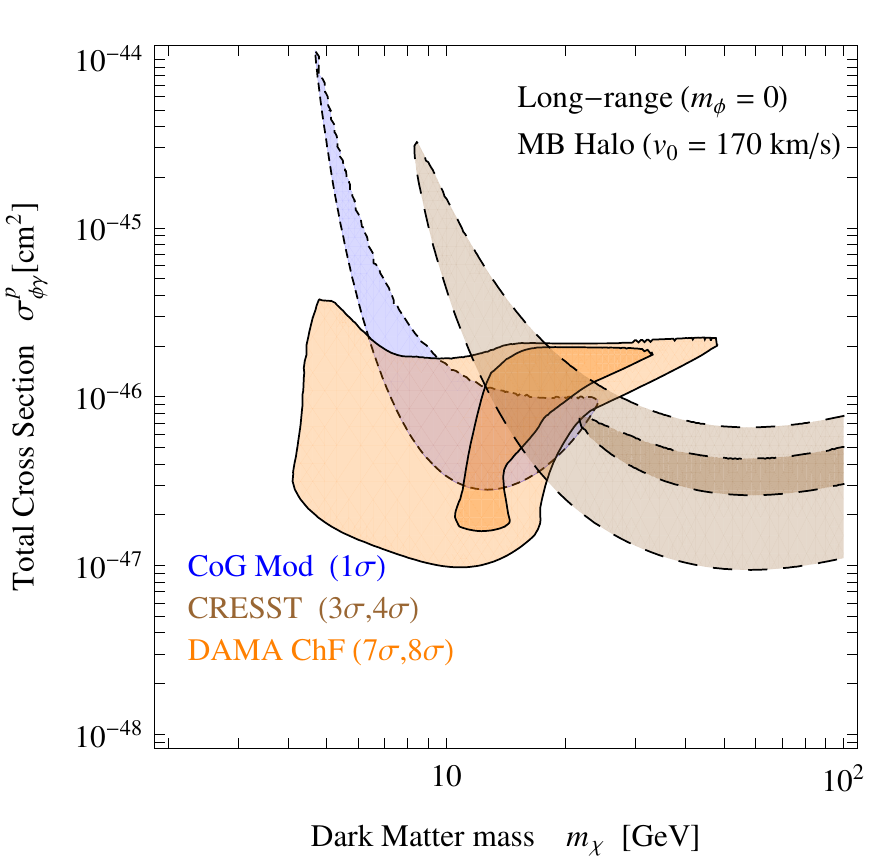}
\includegraphics[width=0.49\textwidth]{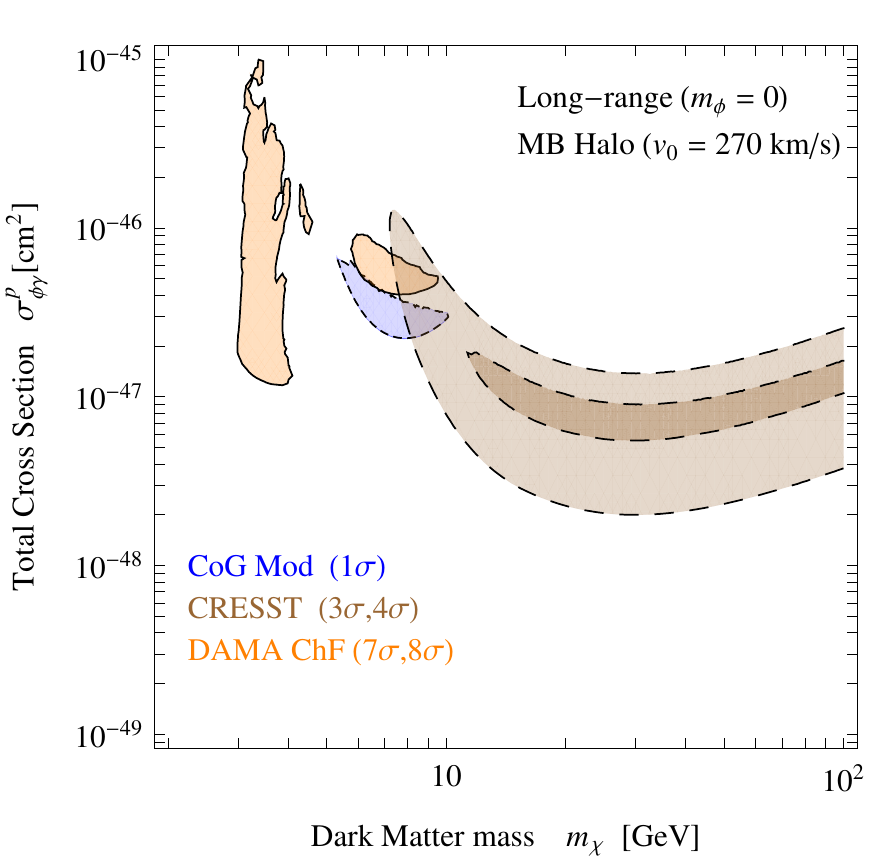}
\caption{The same as in Fig. \ref{fig:C}, except that the interaction is
long--range ($m_\phi = 0$ MeV), for the isothermal sphere and for:
$v_0=170$ km s$^{-1}$ and local density
$\rho_0 = 0.18$ GeV cm$^{-3}$ (left panel);  
$v_0=270$ km s$^{-1}$ and local density
$\rho_0 = 0.45$ GeV cm$^{-3}$ (right panel).
}
\label{fig:Hb}
\end{figure*}

\begin{figure}[t]
\includegraphics[width=0.49\textwidth]{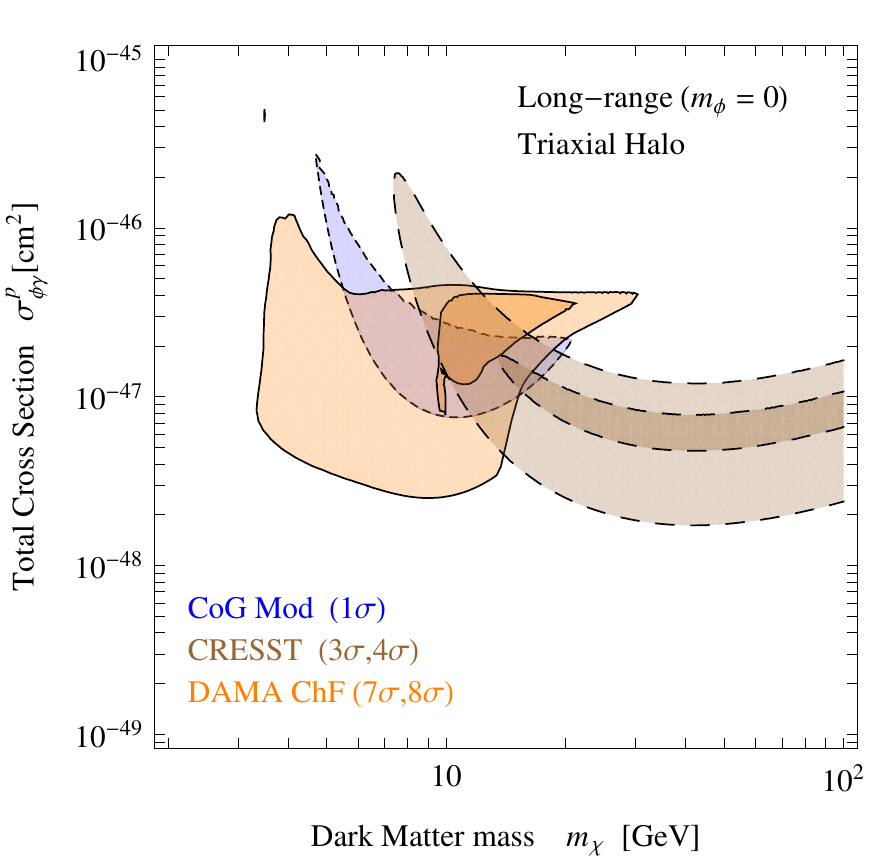}
\caption{The same as in Fig. \ref{fig:C}, except that the interaction is
long--range ($m_\phi = 0$ MeV), for the triaxial halo model. 
}
\label{fig:I}
\end{figure}


\begin{figure*}[t]
\includegraphics[width=0.49\textwidth]{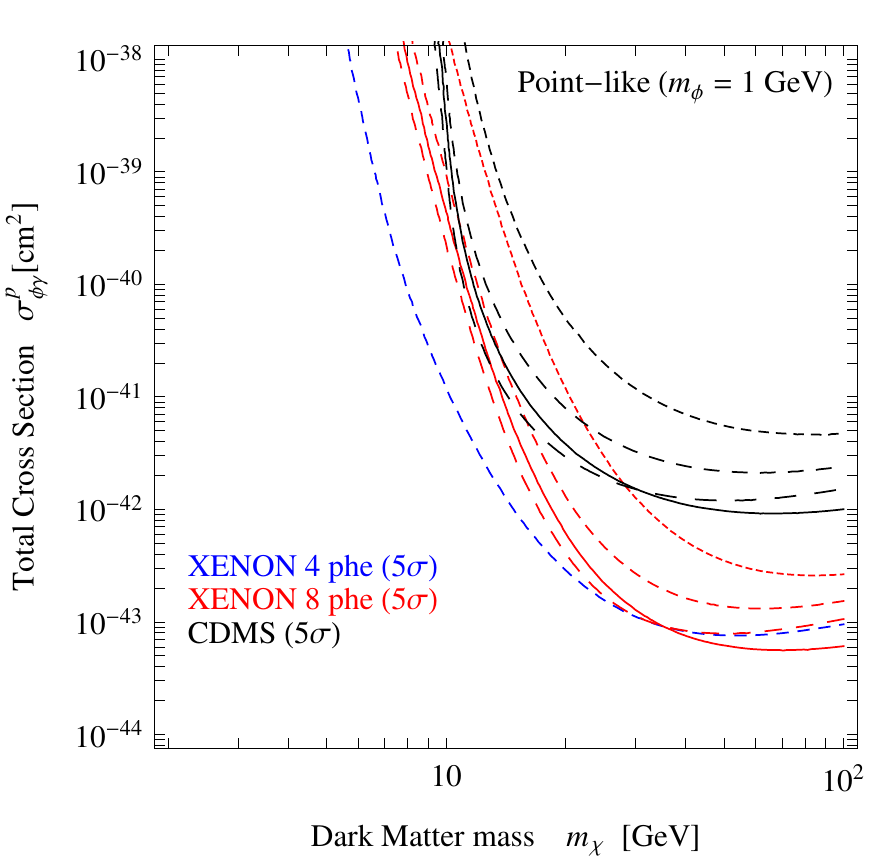}
\includegraphics[width=0.49\textwidth]{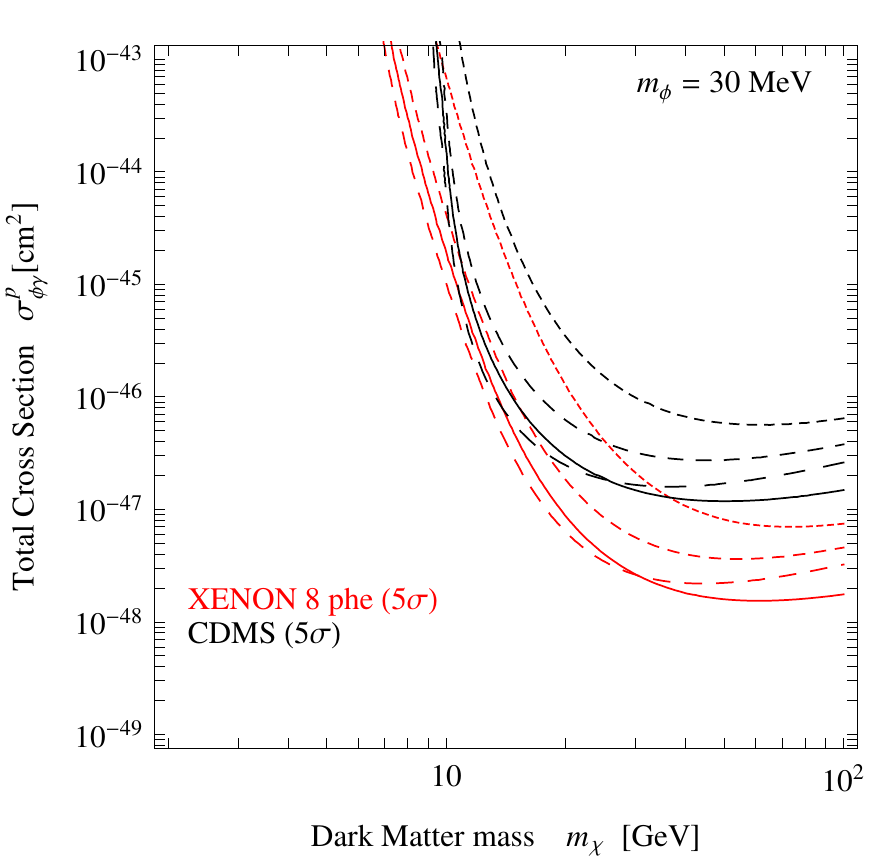}
\includegraphics[width=0.49\textwidth]{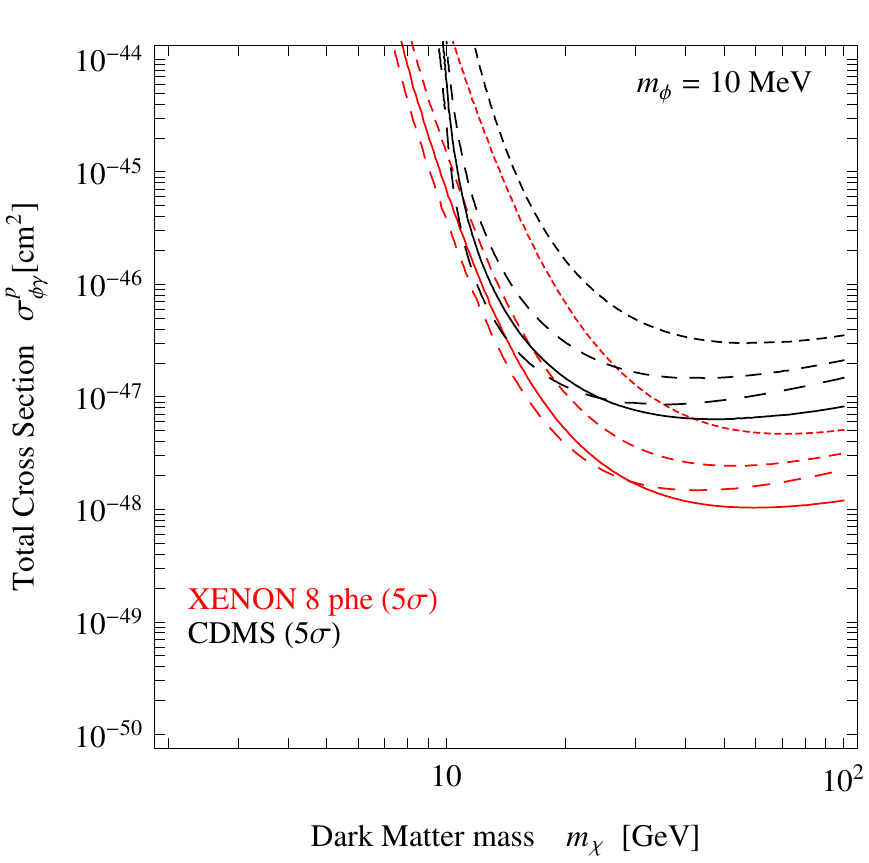}
\includegraphics[width=0.49\textwidth]{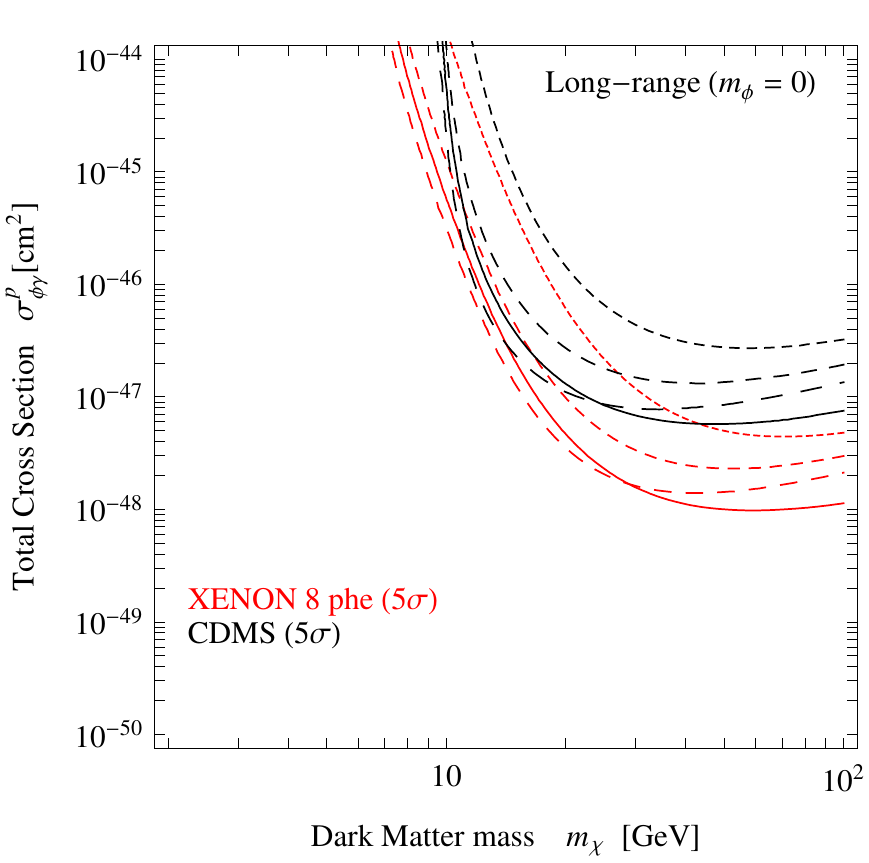}
\caption{Constraints at $5\sigma$ C.L. from CDMS \cite{Ahmed:2009zw} (black lines) and XENON 
\cite{xenon100} (red lines). Broken lines refer to the isothermal sphere with
$v_0 = 170$ km s$^{-1}$ and $\rho_0 = 0.18$ GeV cm$^{-3}$ (short--dashed line),
$v_0 = 220$ km s$^{-1}$ $\rho_0 = 0.3$ GeV cm$^{-3}$ (medium--dashed line),
$v_0 = 270$ km s$^{-1}$ and $\rho_0 = 0.45$ GeV cm$^{-3}$ (long--dashed line). 
Solid lines refer to the triaxial halo
model \cite{Evans:2000gr,Belli:2002yt}. For the XENON detector, all the constraints
refer to a threshold of 8 photoelectron. In the first panel, the blue dashed line
stands for a threshold of 4 photoelectron and for an isothermal sphere with
$v_0 = 220$ km s$^{-1}$ and $\rho_0 = 0.3$ GeV cm$^{-3}$.
}
\label{fig:const}
\end{figure*}

\begin{figure*}[t]
\includegraphics[width=0.49\textwidth]{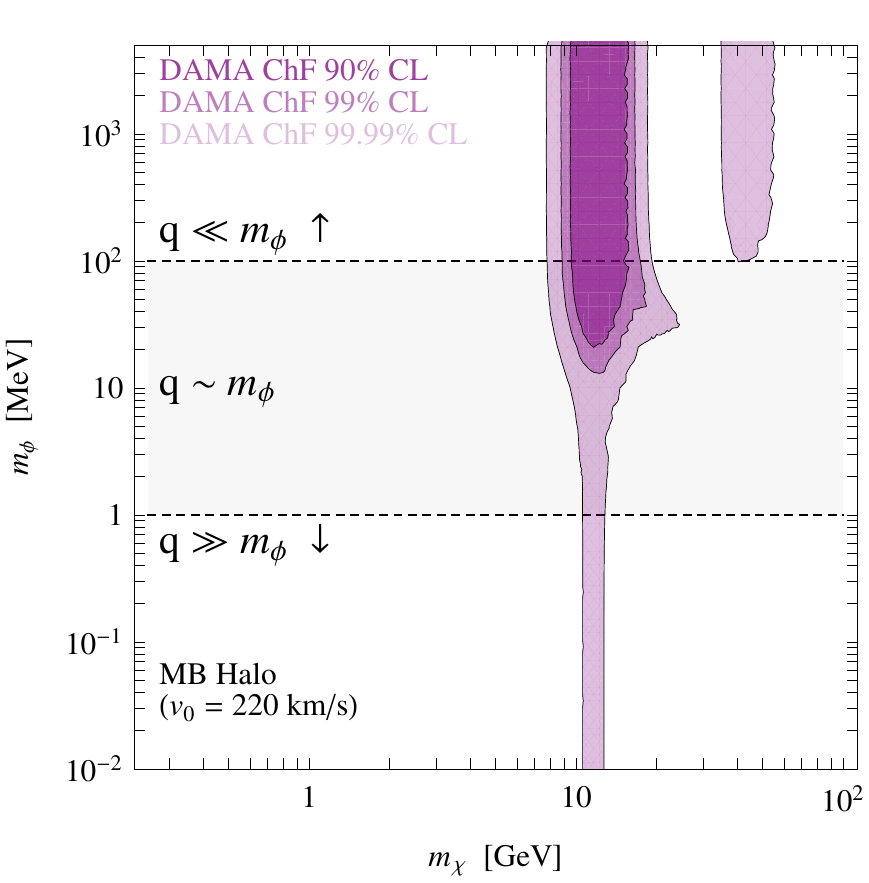}
\includegraphics[width=0.49\textwidth]{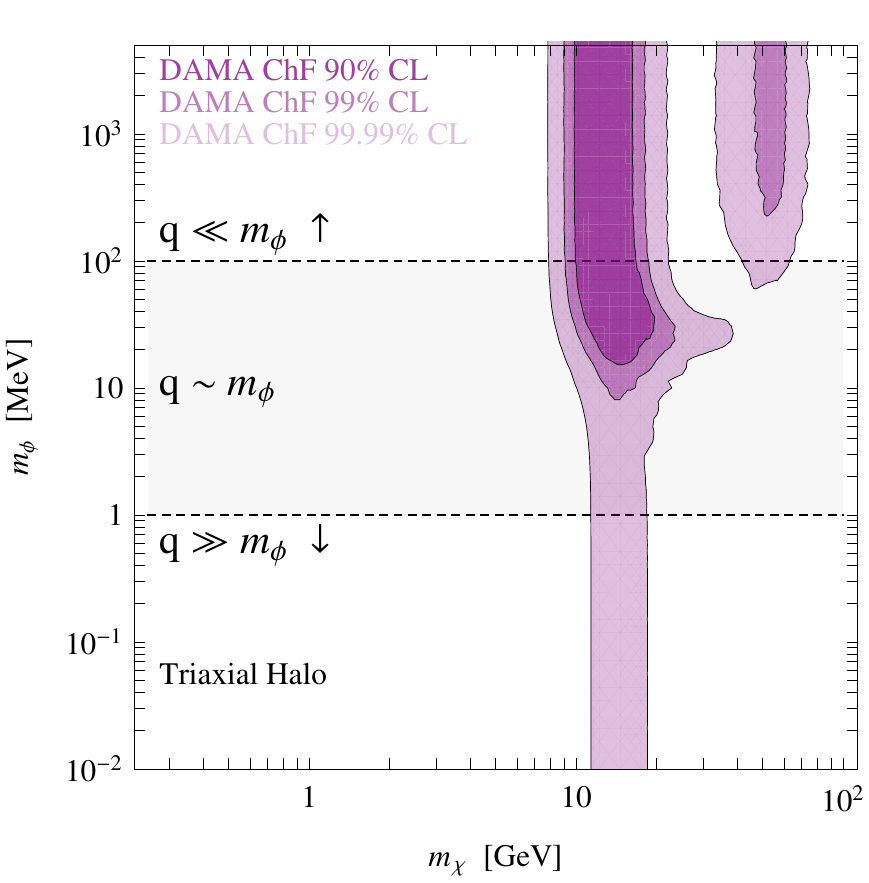}
\caption{Bounds on the light--mediator mass $m_\phi$ as a function of the DM mass
$m_\chi$, as obtained by the analysis of the DAMA annual--modulation result. The
allowed regions are reported for C.L.'s of 90\%, 99\% and 99.99\%.
The left panels stands for an isothermal halo with 
velocity dispersion $v_0=220$ km s$^{-1}$ and local density
$\rho_0 = 0.3$ GeV cm$^{-3}$; the right panels refers to the case of a triaxial
halo with the Earth located on the major axis \cite{Evans:2000gr}, with velocity dispersion 
$v_0=220$ km s$^{-1}$ and local density
$\rho_0 = 0.84$ GeV cm$^{-3}$ \cite{Belli:2002yt}.
}
\label{fig:bounds}
\end{figure*}

\begin{figure}[t]
\includegraphics[width=0.49\textwidth]{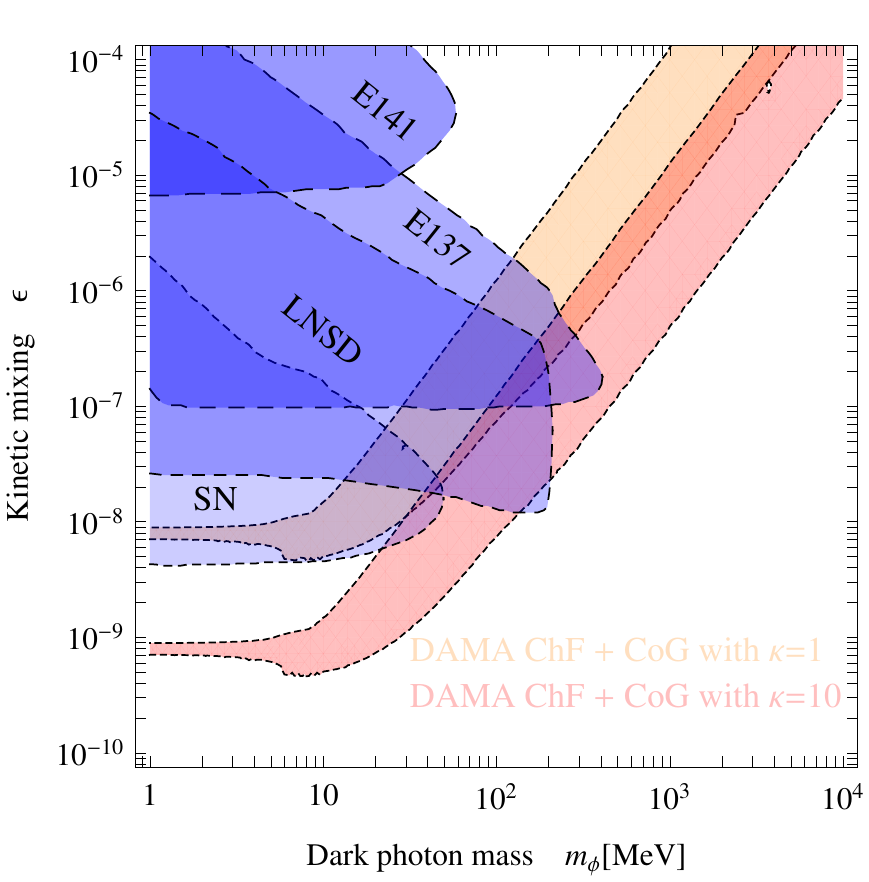}
\caption{ Regions of dark photon mass $m_\phi$ and kinetic mixing $\epsilon$ compatible with DM direct detection experiments, 
as compared with existing independent experimental bounds shown as (blue) areas
on the left (see Refs. \cite{Bjorken:2009mm},\cite{Essig:2010gu} and references therein).  
The upper (orange) and lower (red) regions are derived combining DAMA (with the fraction of channeling marginalized) and CoGeNT annual modulation data, and refer to regions where the absence of a signal can be excluded with a C.L. of 8--$\sigma$. The two regions differ
in the choice of the value of the effective coupling $k$ of dark matter to the dark
photon (defined as $k=Z'\,\sqrt{\alpha_{\rm dark}/\alpha_{SM}}$): the upper (orange)
regions refers to $k=1$, the lower (red) region refers to $k=10$.
}
\label{fig:epsmphi}
\end{figure}


We now proceed to discuss the situation when the mediator mass $m_\phi$ falls in the range
of transition from the point--like to the non point--like case. 
Let us first summarize the effect of long--range interactions in direct detection experiments.
As described in Section \ref{sec:signals}, $d\sigma/dE_R\propto(E_{R}+m_\phi^2/(2m_N))^{-2}$, 
which tells us that larger is the target mass $m_N$ smaller is the recoil energy 
corresponding to the transition from the point--like to the non point--like case, and that,
in the long--range limit, events at low recoil energy are enhanced with respect to the ones
occurring at large recoil energy.
Those simple arguments can be applied to understand the behaviour of the unmodulated signal.
For the modulated signal, on the other hand, the picture is less straightforward.
Indeed, if the minimal velocity providing a nuclear recoil can be significantly smaller than $v_\oplus$
then the (cosine--like) modulated signal is suppressed. 
It can happen at low recoil energy (see Eq. \ref{minvelocity}), and in this case
the long--range interaction would not enhance the modulated signal.

A first set of results
are shown in Fig. \ref{fig:C}, where we report the results for the two reference
values of $m_\phi=10,30$ MeV. The galactic halo is in the form
of an isothermal sphere with local dispersion velocity $v_0=220$ km s$^{-1}$ and
local density $\rho_0 = 0.3$ GeV cm$^{-3}$. The panels on Fig. \ref{fig:C} can
be directly compared to the right panel of Fig. \ref{fig:B}. We notice that moving
from the point--like case to the case of an intermediate--mass mediator, the
allowed regions are sizeably modified (especially for DAMA). 
In the case of DAMA, the DM low--mass region becomes increasingly preferred
over the 100 GeV DM mass region, which basically disappears for $m_\phi=10$ MeV.
The statistical significance of this cases is similar to the point--like case, as can be seen in Tables \ref{tab2} and \ref{tab4}. The same occurs in the case of CoGeNT.
Following the arguments explained above, we find that in the case of scattering on I in DAMA, the unmodulated signal is enhanced at low energy by the long--range interaction while the corresponding modulated signal is not, and so the region at mass above 50 GeV (and most of the channeled region at 10 GeV) is ruled out.
Note also that the enhancement of the signal on Na at very low recoil energies cuts out the region with large cross-section and very small masses.
 In the CRESST case, we consider a signal in O, Ca and W, with the target--mass of the formers being significantly smaller than the latter.
This means that, in the W case, the long--range interaction nature manifests itself at larger mediator mass (or, equivalently, at larger recoil energy).
Therefore, while in the point--like case the region at large DM masses is mostly given by signal on W, the contributions of O and Ca becomes increasingly important as the mediator mass decreases (strictly speaking, when the mass of the mediator decreases, also the cross section needed to reproduce a given signal decreases, until the long--range limit is reached, see Fig. 1, but this process stops earlier in the W case).
In the CoGeNT case, the region remains basically unchanged, partly because we have at disposal only two energy bins, a fact that makes hard any disentanglement of energy--dependent effects.

In order to investigate the dependence of the results on some astrophysical
assumptions, we report in Fig. \ref{fig:D} and \ref{fig:E} the analysis for
intermediate--mass mediator (same reference values: $m_\phi=10,30$ MeV) when the dispersion velocity
of the MB distribution is changed: Fig. \ref{fig:D} refers to $v_0=170$ km s$^{-1}$ 
(which, for consistency in the halo modeling required a
local density $\rho_0 = 0.18$ GeV cm$^{-3}$ \cite{Belli:2002yt}), while Fig, \ref{fig:E}
shows the case of 
$v_0=270$ km s$^{-1}$ 
($\rho_0 = 0.45$ GeV cm$^{-3}$ \cite{Belli:2002yt}). As expected
from kinematics of the DM scattering process, lower average velocities of DM (smaller
values of $v_0$) require larger values of the DM mass in order to reproduce the same
effect in a detector: this implies an extension of the allowed regions toward larger DM
masses for $v_0=170$ km s$^{-1}$. This is manifest in Fig. \ref{fig:D}. The opposite
is true when $v_0$ becomes large, as is clear from Fig. \ref{fig:E}. 
Tables \ref{tab2} and \ref{tab4} show that all these models with intermediate--mass
mediators are viable at the same level as the point--like scenario, with no significant
variation of the statistical significance, except again for the case $v_0=270$ km s$^{-1}$ which gets worse
when the mass of the light mediator becomes small. This is confirmed in Table \ref{tab4},
where the case $m_\phi=0$ is reported. We can conclude that large dispersion velocities
and light mediators are slightly disfavored by the DAMA data (still not at the level of
considering them as excluded), while in the case of dispersion velocities in the range from $v_0=170$ km s$^{-1}$ to about $v_0=220$ -- 250 km s$^{-1}$ the statistical agreement is
basically stable over variation of $m_\phi$ and $v_0$. In the case of CoGeNT, we do not observe
significant variation of the statistical significance when we change $m_\phi$: mild preference toward smaller values of $v_0$ are present also with intermediate--mass mediators, like in the case of point--like interactions, but not statistically significant.

To further discuss the dependence on astrophysics, we extend the analysis to a
different velocity distribution function, by adopting an anisotropic halo model, 
instead of the isotropic MB velocity distribution. For this, we adopt a triaxial
halo model where the Earth is located on the major axis and the velocity anisotropy
is tangential \cite{Evans:2000gr}.
The anisotropic velocity dispersion is taken as in model D2 of Ref. \cite{Belli:2002yt}. The
results for this class of halo models are shown in Figs. \ref{fig:F} and \ref{fig:G}
for the case of a point--like cross section (Fig. \ref{fig:F} ) and in the case of 
mediator masses $m_\phi = 10, 30$ MeV (Fig. \ref{fig:G}). This model presents
a relatively large degree of anisotropy in the velocity dispersion, and this
sizably modify the recoil rate \cite{Belli:2002yt}. When compared to the isotropic MB case,
the triaxial halo model tends to somehow enlarge the range of allowed DM masses, 
especially in the case of intermediate--mass mediator. From the Tables, we notice
that the triaxial halo models are all viable.

The case of very long--range forces is shown in Fig. \ref{fig:Ha} and \ref{fig:Hb}
for the isothermal sphere 
and in Fig. \ref{fig:I} for the triaxial halo model, where we set the extreme case
of $m_\phi=0$, and the statistical significances are reported in Table \ref{tab5}. 
We notice that in the case of $m_\phi=0$, the allowed regions moves toward lighter
DM, even sizably like in the case of an isothermal sphere with large dispersion
velocities. However this last case is disfavored by the statistical analysis
shown in Table \ref{tab5}: while the analysis of the CoGeNT data do not exhibit
significant variation with the change of halo model, in the case of DAMA a very
long--range interaction together with large dispersion velocities is significantly
worse (by about $2\sigma$) than the other cases under analysis.
The CoGeNT, CRESST, and DAMA data can have a common DM interpretation and 
the compatibility increases for large--intermediate $m_\phi$ and small--intermediate $v_0$ 
(see Figs.~\ref{fig:B}--\ref{fig:I}), leading to similar conclusions as the ones derived above for each single datasets.

Constraints that can be derived from the null experiments (CDMS and XENON 100)
are shown in Fig. \ref{fig:const}. The different lines refer to the various
galactic halo models discusses in our analyses: broken lines refer to the isothermal 
sphere with
$v_0 = 170$ km s$^{-1}$ and $\rho_0 = 0.18$ GeV cm$^{-3}$ (short--dashed line),
$v_0 = 220$ km s$^{-1}$ $\rho_0 = 0.3$ GeV cm$^{-3}$ (medium--dashed line),
$v_0 = 270$ km s$^{-1}$ and $\rho_0 = 0.45$ GeV cm$^{-3}$ (long--dashed line). 
Solid lines refer to the triaxial halo
model \cite{Evans:2000gr,Belli:2002yt}. For the XENON detector, all the constraints
refer to a threshold of 8 photoelectron. In the first panel, the blue dashed line
stands, instead, for a threshold of 4 photoelectron and for an isothermal sphere with
$v_0 = 220$ km s$^{-1}$ and $\rho_0 = 0.3$ GeV cm$^{-3}$. We can notice the extent
of variation of the constraints when the galactic halo model and/or the mechanism
of interaction is varied. 
As expected, when comparing long--range interactions with the point--like case, 
the relative impact of the bounds from XENON and CDMS at small DM mass is enhanced with respect to large mass, 
as can be seen by comparing the first panel to the other three.
However, this effect is less pronounced than in experiments with lower threshold (such as DAMA and CoGeNT).

\subsection{Constraints on the mass of the light mediator and on the kinetic mixing}

We finally move to derive bounds on the light--mediator mass $m_\phi$
and on the kinetic mixing parameter $\epsilon$, under the hypothesis
that the mechanism of long--range forces is compatible with the
annual modulation results in direct detection. 

As a first analysis, we discuss the bounds on the mediator mass
by showing the allowed regions
in the plane $m_\phi$ vs. the dark matter mass $m_\chi$, obtained
by adopting a maximum likelihood method. For definiteness, we
adopt the DAMA data set, since this shows a clear modulation effect with
a very large C.L.
The results
are reported in Fig. \ref{fig:bounds}.
The case of
an isothermal sphere with $v_0=220$ km s$^{-1}$ is shown in the left panel,
while the triaxial halo case is reported in the right panel. The regions
are obtained by marginalizing over the channeling fraction.

For both halo cases, two regions of compatibility are present, one for DM
masses around 10 GeV and one for DM masses close to 50--60 GeV. In the case
of heavier DM, long--range forces are excluded (the mass bound
on $m_\phi$ is set around 100 MeV). In the case of light DM, which is
also the overall preferred region, long range forces are viable for the whole range of
the mediator masses. However, the best agreement is obtained for mediator
masses larger than 20 MeV. A 99\% C.L. lower bound on $m_\phi$ is about  10 MeV.
A variation of the DM halo properties does not change dramatically the results,
as long as the (mean) dispersion velocity is not changed. In fact, the comparison
between the isotropic (in velocity space) isothermal sphere and the anisotropic 
triaxial model shows that the allowed region is slightly enlarged in the anisotropic
case: both with respect to the DM mass and to the mass of the mediator $\phi$.
 
 In  Fig. \ref{fig:epsmphi} we show the bounds obtained
in the plane $\epsilon$ -- $m_\phi$. In this case, we choose to focus 
on bounds derived for a kinetic mixing between dark and ordinary photons,
which is described by a lagrangian term of the form $\frac{\epsilon}{2}\,F_{\mu\nu}^{SM} F'^{\mu\nu}_{\rm dark}$. These type of couplings can produce experimentally observable effects in many different physical situations. A summary of those constraints are
discussed e.g. in Refs. \cite{Bjorken:2009mm}, \cite{Essig:2010gu} (and references therein) and are reproduced in Fig. \ref{fig:epsmphi} as (blue) regions on the left. They constrain large couplings and very light mediators.

Coming to the relevant regions which can be derived from the analysis of DM direct detection
experiments, we first notice that the DM signal depends on four `dark sector' parameters: $m_\phi$, $\epsilon$, $Z'$, and $\alpha_{\rm dark}$.
Essentially, it involves an extra coupling (between DM and dark photon, proportional
to $Z'\sqrt{\alpha_{\rm dark}}$) with respect to dark photon signals in laboratory experiments.
Therefore, there is more freedom associated, and observational bounds on the dark photon properties can only partially constrain the allowed parameter space.
In this respect, we define $k=Z'\,\sqrt{\alpha_{\rm dark}/\alpha_{SM}}$ and consider two benchmark cases, $k=1$ and $k=10$.
In Fig. \ref{fig:epsmphi}, we show a combined fit to DAMA and CoGeNT annual modulation data. The analysis is performed along the line explained in Sec. \ref{sec:analysis}.
We combine the two data sets by adding the $\chi^2$ of the two experiments together, and show those regions where the absence of signal can be excluded at 8--$\sigma$  C.L.. For DAMA, we vary the fraction of channeling in its allowed range. 
From Fig. \ref{fig:epsmphi} we see that in the
`totally symmetric' case with $k=1$ (upper orange region), light mediators are excluded, and only dark photons with $m_\phi>100$ MeV can simultaneously satisfy the constraints and provide a suitable interpretation for DAMA and CoGeNT data. However, for $k\gtrsim10$, the whole range of light--mediator masses is allowed. It is worth to point out that such values of $k$ can be easily predicted in models with a strongly coupled dark sector or including `composite DM' with large dark charge.

Another class of constraints arises from bounds on DM self--interaction in cosmic structures, a bound which depends on the DM coupling with the dark photon and on the mass of the dark photon, while it is independent on the kinetic mixing parameter $\epsilon$.
The long range force between DM particles implies that DM is more collisional than in the standard WIMP case.
A first example of such constraints comes from observations of systems of colliding galaxy clusters, such as the Bullet cluster~\cite{Clowe:2006eq}, which point towards collisionless DM.
A robust bound of $\sigma/m_\chi \leq 1.25$ cm$^2$/g has been placed on the size of the self-scattering of DM~\cite{Randall:2007ph}. 
To infer the constraint on the mass of the dark photon, we compute the weighted cross section:
\begin{equation}
\sigma_{\rm av}=\int d^3 v_1 d^3 v_2 \, f(v_1) f(v_2)\int d\Omega \frac{d\sigma}{d\Omega}(1-\cos\theta)
\end{equation}
that measures the rate at which energy is transferred in the system of colliding clusters. Here $f(v)=(\pi v_0^2)^{3/2}\,e^{-(v/v_0)^2}$ is the velocity distribution of dark matter particles  in the cluster, assumed to be Maxwellian, $v_0$ is the velocity dispersion, $\theta$ is the scatter angle in the centre of mass frame and the cross section
on the right hand side refers to the elastic processes. For dark matter particles interacting with long range forces, the self-interaction is analogous to a screened Coulomb scattering with the plasma, which is well fit by the following cross section~\cite{Feng:2009hw,Loeb:2010gj}:
\begin{equation}
\sigma(v_{\rm rel})\simeq 
\left\{\begin{array}{lc}
\displaystyle \frac{4\pi}{m_\phi^2} \beta^2 \ln\left(1+\beta^{-1}\right), & \beta\lesssim 0.1, \\
\displaystyle \frac{8\pi}{m_\phi^2} \beta^2 /\left(1+1.5 \,\beta^{1.65} \right), & 0.1\lesssim\beta\lesssim 10^3, \\
\displaystyle \frac\pi{m_\phi^2}\left(1+\ln\beta-0.5\,\ln^{-1}\beta\right)^2, & \beta\gtrsim 10^3,
\end{array}\right.
\end{equation}
where $\beta=2\,Z'^2\,\alpha_{\rm dark} m_\phi/(m_\chi v_{\rm rel}^2)$ and $v_{\rm rel}=|\vec v_1-\vec v_2|$ is the relative velocity of dark matter particles. Considering typical velocity of collision in the Bullet cluster of 4700 km/s, 
a dark matter mass of 10 GeV (to fit direct detection observations) and two values of the parameter $k=(1,10) \Rightarrow Z'^2\,\alpha_{\rm dark}=(\alpha_{\rm SM},100\,\alpha_{\rm SM})$, we get that the bound on the self--interaction is exceeded if the mass of the dark photon is smaller than (1, 20) MeV.

A second class of constraints comes from the fact that 
large DM self--interactions causes a rapid energy transfer between DM particles and thus tends to drive DM halos into spherical isothermal configurations~\cite{Feng:2009hw,Feng:2009mn,Buckley:2009in,Ibe:2009mk,Loeb:2010gj}.
The observation of DM halo ellipticity allows to put further constraints on the size of self interaction. 
To estimate the impact that self interaction via screened Coulomb scattering has in isotropizing the shape of dark matter halo, we compute the relaxation time $\tau_{\rm r}=\Gamma_{\rm av}^{-1}$ where 
\begin{equation}
\Gamma_{\rm av}=\int d^3 v_1 d^3 v_2 \,f(v_1)f(v_2)\,n_\chi v_{\rm rel} \sigma(v_{\rm rel}) \left( v_{\rm rel}/v_0\right)^2,
\label{eq:relax}
\end{equation}
is the rate at which energy is transferred in the system. This relaxation time provides an estimate on the effects of self interactions on the  dynamics of a virialized astrophysical object with number density $n_\chi=\rho_\chi/m_\chi$ and velocity dispersion $v_0$. Indeed if it is much longer than the age of the object (say $\tau\sim 10^{10}$ years for ``old'' objects), we expect that self interaction does not alter the dynamics. On the other hand, much shorter relaxation times imply that an isothermal configuration tends to form and thus such scenario would be excluded by the observation of few elliptical halos~\cite{Feng:2009hw,Feng:2009mn,Buckley:2009in,Ibe:2009mk,Loeb:2010gj}. 
In the case of galaxy clusters ($\rho_\chi\sim 10^{-5}$ GeV/cm$^3$ and $v_0\sim1000$ km/s), considering again $m_\chi=10$ GeV and $k=(1,10)$, we get that the relaxation time is always much longer than the age of clusters and thus the mass of the dark photon is basically unbounded. 

On the other hand, following the same analytic approach for virialized halos at galactic scales,  where the dark matter energy density is higher ($\rho_\chi \sim 1$ GeV/cm$^3$) and velocity dispersion is lower ($v_0\simeq 240$ km/s for a galaxy and $v_0\simeq 10$ km/s for a dwarf galaxy), one can in principle put very stringent constraints on $m_\phi$, excluding $m_{\phi}\lesssim100$ MeV. 
However, Eq.~(\ref{eq:relax}) is only an approximation, and since galaxies form in a fully non--linear regime, an N--body numerical simulation explicitly including long--range interactions would be in order to properly address the size of the constraint. Moreover, we note that interactions with baryons (possibly in the long--range regime as well) could substantially affect the halo shape. This very interesting subject deserves a dedicate treatment which is beyond the goal of this paper.

\section{Conclusions}

In this paper we have discussed the current 
positive indications of a possible dark matter signal in direct detection experiments, in terms of a mechanism of interaction between the dark matter particle and the
nuclei occurring via the exchange of a light mediator, resulting in a long--range interaction. 
We have therefore analyzed the annual modulation results observed by the DAMA and CoGeNT experiments under the hypothesis of this type of long--range interactions, and derived
bounds on the relevant parameters at hand, namely the DM mass and the effective cross section,
as well as the mass of the light mediator. 

We find that long--range forces are a viable mechanism which is able to explain the modulation
effects, and we have obtained that the preferred range of masses for the light mediator 
is for values larger than a few MeV. This is correlated with a DM mass around 10 GeV. We have
also obtained that the long--range forces mechanism is more constrained in the case of
large dispersion velocities of the DM particles in the galactic halo: in the case
of dispersion velocities larger than about $v_0 \sim 250$ km s$^{-1}$ the long--range forces
are less favored (although not excluded). This result is stable over the whole variation
of the light--mediator mass. We can therefore conclude that the annual modulation effect reported by DAMA is compatible with long--range forces when the DM mass is preferably in the range from 8 GeV to 20 GeV and the mediator mass is in excess of about 10 MeV, depending
on the actual galactic halo model.

\acknowledgments

We would like to thank P. Belli, R. Bernabei and J. Collar for many useful and
interesting discussions on experimental and analysis issues and P. Serpico for
exchanges on theoretical topics.
We acknowledge research grants funded jointly by Ministero
dell'Istruzione, dell'Universit\`a e della Ricerca (MIUR), by
Universit\`a di Torino and by Istituto Nazionale di Fisica Nucleare
within the {\sl Astroparticle Physics Project} (MIUR contract number: PRIN 2008NR3EBK;
INFN grant code: FA51). P.P. acknowledges support from Consorzio Interuniversitario per la Fisica Spaziale (CIFS) under contract number 33/2010.
N.F. acknowledges support of the spanish MICINN
Consolider Ingenio 2010 Programme under grant MULTIDARK CSD2009- 00064.
This work was partly completed at the Theory Division of CERN in the context of the TH--Institute `Dark Matter Underground and in the Heavens' (DMUH11, 18-29 July 2011).

\end{document}